\documentclass[aps,preprint,amsmath,amssymb,amsfonts]{revtex4}
\usepackage{epsfig}
\usepackage{graphicx}
\usepackage{subfigure}
\usepackage{dcolumn}
\usepackage{bm}
\usepackage{amsthm}
\usepackage{enumitem}
\usepackage{slashed}
\usepackage{braket}
\usepackage{amsmath}
\usepackage{mathtools}
\usepackage{bbold}

\usepackage{color}   
\usepackage{hyperref}
\hypersetup{
	colorlinks=true,  
	linkcolor=black,  
	citecolor=black,
	filecolor=black,
	urlcolor=black,
}

\makeatletter
\def\l@subsubsection#1#2{}
\makeatother

\newcommand{\const}{\mbox{const}}
\newcommand{\del}{\partial}

\renewcommand{\Im}{\operatorname{Im}}
\newcommand{\sign}{\operatorname{sign}}

\newcommand{\bbP}{\mathbb{P}}
\newcommand{\bbR}{\mathbb{R}}

\newcommand{\calD}{\mathcal{D}}

\newcommand{\calG}{\mathcal{G}}

\newcommand{\calJ}{\mathcal{J}}
\newcommand{\calK}{\mathcal{K}}
\newcommand{\calL}{\mathcal{L}}
\newcommand{\calM}{\mathcal{M}}

\newcommand{\calP}{\mathcal{P}}

\newcommand{\calR}{\mathcal{R}}

\newcommand{\calV}{\mathcal{V}}
\newcommand{\scri}{\mathcal{I}}

\newcommand{\dSCFT}[2]{dS\textsubscript{#1}/CFT\textsubscript{#2}}

\begin{document}

\title{Causality of higher-spin interactions on the (A)dS lightcone, with application to the static patch}

\author{Jin Kozaki}
\email{jin.kozaki53@gmail.com}
\affiliation{Okinawa Institute of Science and Technology, 1919-1 Tancha, Onna-son, Okinawa 904-0495, Japan}

\author{Julian Lang}
\email{julian.lang.research@gmail.com}
\affiliation{Okinawa Institute of Science and Technology, 1919-1 Tancha, Onna-son, Okinawa 904-0495, Japan}

\author{Yasha Neiman}
\email{yashula@icloud.com}
\affiliation{Okinawa Institute of Science and Technology, 1919-1 Tancha, Onna-son, Okinawa 904-0495, Japan}

\date{\today}

\begin{abstract}
We study Higher-Spin Gravity in 4-dimensional (Anti-)de Sitter space, at leading order in the interactions (cubic vertices), in the AdS lightcone formalism developed by Metsaev. Using the vertices' chiral structure, we extend the formalism into a broader class of lightcone frames, which allows for lightcones of bulk points. This enables us to write the lightcone theory in de Sitter space, where only these more general frames are available. It also allows us to formulate and verify (for the first time!) some causal properties of massless higher-spin interactions, involving lightcone foliations that share a lightray. These causal properties serve to both motivate and enable the computation of ``static-patch scattering amplitudes'' -- the evolution of fields between the two horizons of the maximal observable region in de Sitter space. We present a computation scheme for such ``amplitudes'' in coordinate space, and in momentum space with spinor-helicity variables.
\end{abstract}

\maketitle
\tableofcontents
\newpage

\section{Introduction} \label{sec:intro}

\subsection{Scope and goals}

Higher-Spin (HS) Gravity is the putative interacting theory of an infinite tower of massless gauge fields with increasing spin. It can be thought of as a smaller cousin of string theory, with stringy features such as an AdS/CFT holographic formulation \cite{Klebanov:2002ja,Sezgin:2002rt,Sezgin:2003pt,Giombi:2012ms}, but ``native'' to 4 spacetime dimensions. Here, we will focus on the so-called minimal type-A theory, which has a single parity-even field of every even spin $s=0,2,4,\dots$. What is remarkable about this version of the theory is that its AdS/CFT duality can be extended to de Sitter space, providing a working model of \dSCFT{4}{3} \cite{Anninos:2011ui}. The grain of salt is that this simplest version of HS theory does not have a limit where the higher-spin fields decouple, or where the graviton's interactions are those of General Relativity (GR). 

This paper is part of a research program to explore HS theory as a working model of quantum gravity in $dS_4$. Specifically, we work on the ``bulk side'' of this program, where we aim to compute observables in the maximal causal region of de Sitter space -- the static patch -- directly in the bulk theory, with no reference to holography. Our chosen observables are ``scattering amplitudes'' -- more properly, the functional dependence of final field data on initial field data -- between the past and future boundaries of the static patch, i.e. between the past and future cosmological horizons of a de Sitter observer. These were explored at the free level for massless fields of all spins in \cite{David:2019mos}, and for interacting lower-spin fields (up to Self-Dual GR) in \cite{Albrychiewicz:2020ruh,Albrychiewicz:2021ndv,Neiman:2023bkq}. Some of the relevant steps were also extended \cite{Neiman:2024vit,Lang:2025rxt} to higher-spin fields with lower-spin (1-derivative and 2-derivative, or Yang-Mills-like and GR-like) interactions. Our general method for computing static-patch scattering can be described as follows:
\begin{enumerate}
	\item We work in a Poincare coordinate patch, in a lightcone gauge \footnote{For an application of lightcone gauge to the more standard problem of AdS boundary correlators, see \cite{Chowdhury:2024dcy}.}. The initial data on the past horizon is then expressed as data on ``past lightlike infinity'' in these Poincare coordinates, or, equivalently, as massless plane-wave modes.
	\item We evolve this initial data into the bulk of the Poincare patch, until we reach one of the lightrays of the \emph{future} horizon (specifically, the lightray pointing along the preferred direction of our lightcone gauge). 
	\item We transform the final data on this lightray from the original lightcone gauge into the one adapted to the future horizon.
	\item Repeat for every choice of preferred lightlike direction, so as to obtain the final data on every lightray of the future horizon.
\end{enumerate}
In the present paper, we aim to extend this method to ``true'' higher-spin interactions, i.e. interactions with arbitrary numbers of derivatives, acting on fields with arbitrary spin. We will focus on the lowest order in interactions, i.e. on the cubic vertices, which are well-known in several formalisms -- see \cite{Sleight:2016dba} for the vertices in the language of Fronsdal fields \cite{Fronsdal:1978rb,Fronsdal:1978vb}, or \cite{Gelfond:2018vmi,Didenko:2018fgx,Didenko:2019xzz,Gelfond:2019tac} for their derivation from Vasiliev's equations \cite{Vasiliev:1990en,Vasiliev:1995dn,Vasiliev:1999ba}. In this paper, we will instead use the lightcone formulation, given in \cite{Ponomarev:2016lrm,Skvortsov:2018uru}, using the framework developed in \cite{Bengtsson:1983pd,Fradkin:1991iy,Metsaev:1999ui,Metsaev:2003cu,Metsaev:2018xip}. This is in keeping with the strategy outlined above for the static-patch scattering problem: a lightcone formalism naturally reduces the gauge-redundant covariant fields to their physical degrees of freedom, in a frame adapted to a static patch's lightlike horizons, where initial/final data can be encoded. The main difference between the present paper and our previous work on lower-spin interactions \cite{Albrychiewicz:2021ndv,Neiman:2023bkq,Neiman:2024vit,Lang:2025rxt} is that in the latter we started from a covariant formulation, and then adopted a lightcone \emph{gauge}. For general higher-spin interactions, that path remains possible, but seems difficult. Thus, in this paper, we'll skip the covariant formulation, and instead start directly from a lightcone formalism. Having made that choice, two difficulties arise:
\begin{enumerate}
	\item At the cubic order, HS theory makes sense for any value of the cosmological constant $\Lambda$ (with an expectation that $\Lambda\neq 0$ becomes necessary at higher orders). However, the lightcone formulation has been given only for $\Lambda\leq 0$. Before we can use the lightcone formalism in de Sitter space, we'll need to extend it appropriately.
	\item The motivation for our static-patch calculation is the causal structure of de Sitter space. However, very little is known about causality in massless higher-spin interactions, or, for that matter, in the lightcone formalism. We'll need to formulate and demonstrate the relevant causality properties.
\end{enumerate}
Fortunately, it seems that both of these difficulties have the same solution! Let us start with the problem of adapting the AdS lightcone formalism \cite{Metsaev:1999ui,Metsaev:2003cu,Metsaev:2018xip} to de Sitter space. The specific obstacle here is that in \cite{Metsaev:1999ui,Metsaev:2003cu,Metsaev:2018xip}, the preferred lightlike direction $x^-$ is always orthogonal to the AdS warp factor $z$. In de Sitter, this restriction cannot be satisfied, since $z$ is replaced by a timelike coordinate $t$. We're thus led to develop a more general class of lightcone frames, which allows for non-orthogonal $x^-$ and $z$. Geometrically, this means allowing foliations of spacetime into \emph{lightcones of bulk points}, as opposed to lightcones of boundary points. This generalization allows us to ask new questions about causality, which were not available in the original lightcone formalism! In particular, we can consider two lightcone frames involving \emph{two different lightcones that share a lightray}, and ask whether this ray gets mapped to itself in the frame transformation. As we will see, the answer is yes, and it's of direct relevance to the static-patch scattering problem.

Throughout, we'll make extensive use of the helicity/chirality structure of massless fields in 4d and their interaction vertices. In particular, the cubic vertices in the lightcone formalism decompose into two sectors -- chiral and anti-chiral -- related by complex conjugation. By itself, each of these sectors generates a self-contained, though non-unitary, HS theory \cite{Skvortsov:2018jea,Skvortsov:2020wtf,Skvortsov:2022syz,Sharapov:2022faa,Sharapov:2022awp,Didenko:2022qga} (at $\Lambda=0$ in the lightcone formulation, the chiral theory is complete without higher-order vertices; at $\Lambda\neq 0$, this is an open question). Our key result is that, in the chiral sector, the generalized lightcone formalism can be read off from the original one 
\cite{Metsaev:2018xip} via an \emph{analytic continuation}. Thus, we will first restrict to the chiral theory, then analytically continue, and then return to the real theory by adding in the anti-chiral complex conjugate.

Though our main focus is on the minimal theory with even spins, we will occasionally discuss Yang-Mills-like interactions with total helicity $\pm 1$, as the simplest example of massless interactions with a chirality structure. To be non-trivial, such interactions require the fields to carry color factors, which we will omit.

\subsection{Summary and structure of the paper}

The paper is structured as follows. In section \ref{sec:framework}, we introduce Metsaev's AdS lightcone formalism for cubic-level HS theory. We do this in a more covariant notation than the original paper \cite{Metsaev:2018xip}, while casting the AdS isometry group as sitting inside the larger (but broken) conformal group. This turns out to be useful, because some conformal generators, even though broken, nevertheless appear as building blocks inside the unbroken isometry generators. Some special cases where the conformal group is unbroken, i.e. free fields and Yang-Mills-like interactions, are analyzed in Appendix \ref{app:conformal}. In section \ref{sec:extend:chiral}, we focus on the chiral theory, and shift into a chiral field frame adapted to it. In section \ref{sec:extend:analytic_continuation}, we capitalize on all this rewriting by extending the lightcone formalism to foliations with bulk lightcones, and to de Sitter space. In section \ref{sec:extend:real}, we re-introduce the anti-chiral sector. In section \ref{sec:geometry}, we expand on the geometric meaning of our generalized lightcone formalism. In section \ref{sec:causality}, we describe the new causality properties that the generalized lightcone formalism makes apparent. As a side benefit, we point out in section \ref{sec:causality:covariant} how the lightcone fields can be packaged into more covariant quantities (which, at the linearized level, coincide with Weyl curvature components). In section \ref{sec:static_patch}, we apply all of the above to set up scattering computations in the de Sitter static patch. As an explicit example, in section \ref{sec:static_patch:GR_like} we consider gravity-like interactions with total helicity $h_1+h_2+h_3=2$. This includes Self-Dual GR and its HS generalization \cite{Krasnov:2021nsq}, previously discussed in this context in \cite{Neiman:2023bkq,Neiman:2024vit,Lang:2025rxt}. Section \ref{sec:discuss} is devoted to discussion and outlook. In Appendix \ref{app:matching}, we patch a small gap in the literature, by mapping the cubic coupling constants between the lightcone formalism and the language of Fronsdal fields (where the couplings are fixed by holography).

\section{Review and rewriting of cubic HS theory on the lightcone} \label{sec:framework}

\subsection{Coordinates and indices}

We work in 4d spacetime with coordinates $x^\mu$, with a metric of mostly-plus signature. We will mostly use Poincare coordinates, raising and lowering their indices with the Minkowski metric $\eta_{\mu\nu}$. The actual spacetime metric is either $\eta_{\mu\nu}$, $\eta_{\mu\nu}/z^2$ or $\eta_{\mu\nu}/t^2$, for Minkowski, AdS and de Sitter respectively. Here, $z$ and $t$ are coordinates with a constant unit spacelike/timelike gradient respectively: 
\begin{align}
  \del_\mu z\del^\mu z = 1 \ ; \quad \del_\mu t\del^\mu t = -1 \ ; \quad \del_\mu\del_\nu z = \del_\mu\del_\nu t = 0 \ .
\end{align}
We will also use spinor indices, left-handed $(\alpha,\beta,\dots)$ and right-handed $(\dot\alpha,\dot\beta,\dots)$. These are raised/lowered with the flat antisymmetric ``metrics'' $\epsilon_{\alpha\beta},\epsilon_{\dot\alpha\dot\beta}$, according to:
\begin{align}
	\zeta_\alpha = \epsilon_{\alpha\beta}\zeta^\beta \ ; \quad \zeta^\alpha = \zeta_\beta\epsilon^{\beta\alpha} \ ; \quad 
	\bar\zeta_{\dot\alpha} = \epsilon_{\dot\alpha\dot\beta}\bar\zeta^{\dot\beta} \ ; \quad \bar\zeta^{\dot\alpha} = \bar\zeta_{\dot\beta}\epsilon^{\dot\beta\dot\alpha} \ .
\end{align}
We translate between spinor and vector indices via the flat Pauli matrices $\sigma_\mu^{\alpha\dot\alpha}$, as:
\begin{align}
	\xi^{\alpha\dot\alpha} = \sigma_\mu^{\alpha\dot\alpha}\xi^\mu \ ; \quad \xi^\mu = -\frac{1}{2}\sigma^\mu_{\alpha\dot\alpha}\xi^{\alpha\dot\alpha}  \ .
\end{align}
This applies in particular to the spacetime gradient: $\del_{\alpha\dot\alpha} = \sigma^\mu_{\alpha\dot\alpha}\del_\mu$. Rank-2 antisymmetric spacetime tensors are separated into left-handed and right-handed parts as:
\begin{align}
	F^{\alpha\dot\alpha\beta\dot\beta} = F^{\alpha\beta}\epsilon^{\dot\alpha\dot\beta} + F^{\dot\alpha\dot\beta}\epsilon^{\alpha\beta} \ .
\end{align}
In the lightcone formalism, we introduce a preferred constant lightlike vector $\ell^\mu$, with spinor square root $q^\alpha,\bar q^{\dot\alpha}$:
\begin{align}
	\ell^{\alpha\dot\alpha} = 2q^\alpha\bar q^{\dot\alpha} \ ; \quad \del_\mu q^\alpha = \del_\mu\bar q^{\dot\alpha} = 0 \ .
\end{align}
The ``equal-time'' spacetime slices in the lightcone formalism are the null hyperplanes $\ell_\mu x^\mu = \const$. We define an integration measure $d^3x$ and its associated delta-function $\delta^3(x)$ on the hyperplane $\ell_\mu x^\mu = \const$ via:
\begin{align}
	d^3x = d^4x\,\delta(\ell_\mu x^\mu - \const) \ .
\end{align}
In Metsaev's lightcone formalism, it is important that $\ell^\mu$ is chosen orthogonal to the AdS warp factor $z$:
\begin{align}
	\ell^\mu\del_\mu z = 0 \ . \label{eq:ell_z_ortho}
\end{align}
One of our main goals will be to remove this condition, since it cannot be satisfied for the dS warp factor $t$. In terms of the spinors $q^\alpha,\bar q^{\dot\alpha}$, the condition \eqref{eq:ell_z_ortho} can be written as:
\begin{align}
 \bar q^{\dot\alpha} = q^\alpha\del_\alpha{}^{\dot\alpha}z \ , \label{eq:q_bar}
\end{align}
for a certain choice of the complex phase of $q^\alpha,\bar q^{\dot\alpha}$. 

\subsection{Lightcone formalism}

The main strength of the lightcone formalism is that it avoids gauge-redundant tensor fields. Instead, massless particles are described by just one scalar field $\Phi_h(x^\mu)$ for each helicity $h$, with the reality condition $\Phi_{-h} = \Phi_h^\dagger$. The action takes the form:
\begin{gather}
	S =  \int d^4x\,\calL \ ; \quad \calL = \calL_{[2]} + \calL_{[3]} + \dots \ ; \quad \calL_{[2]} = \frac{1}{2}\sum_h \Phi_{-h}\Box\Phi_h \ . \label{eq:action}
\end{gather}
Here, $d^4x$ is the flat 4-volume measure, $\Box \equiv \del_\mu\del^\mu$ is the flat d'Alembertian, and the subscripts in $\calL_{[n]}$ refer to terms of $n$'th order in the fields. The interaction terms $(\calL_{[3]},\dots)$ contain derivatives only along the $\ell_\mu x^\mu = \const$ hyperplane: the transverse derivative only appears inside the $\Box$ in the kinetic term $\calL_{[2]}$. On the $\ell_\mu x^\mu = 0$ hyperplane, the fields satisfy canonical commutation relations:
\begin{align}
 [\Phi_h(x),\Phi_{h'}(x')] = \frac{\delta_{-h,h'}}{2i(\ell\cdot\del)}\,\delta^3(x - x') \ . \label{eq:commutators}
\end{align}
Here, the spatial parts of the delta-function ensure that $x,x'$ are on the same lightray. The delta-function's lightlike part, acted on by the inverse of the lightlike derivative $\ell\cdot\del\equiv\ell^\mu\del_\mu$, produces a sign function that depends on the ordering of $x,x'$ along the lightray. This non-locality is the usual price of the lightcone formalism. As we'll discuss in sections \ref{sec:geometry}-\ref{sec:causality}, there are some ways around it, especially in the cubic interactions' chiral sector. 

Another price for working with just physical degrees of freedom is that the action's spacetime symmetry is not manifest. Thus, we must separately define the spacetime symmetry generators and verify their commutator algebra. We will start by considering full 4d conformal symmetry, which will later be reduced to a Poincare/AdS/dS subgroup. The generators can be written as integrals over the $\ell_\mu x^\mu = \const$ hyperplane, as follows:
\begin{align}
	&\text{Translations:} & P^\mu &= \int d^3x\,\calP^\mu \ ; & \calP^\mu &= \sum_h \Phi_{-h}(\ell\cdot\del)\del^\mu\Phi_h - \ell^\mu\calL \ ; \label{eq:P_general} \\
	&\text{Lorentz:} & J^{\mu\nu} &=  \int d^3x\,\calJ^{\mu\nu} \ ; & \calJ^{\mu\nu} &= 2x^{[\mu}\calP^{\nu]} + \calM^{\mu\nu} \ ; \label{eq:J_general} \\
	&\text{Dilatations:} & D &=  \int d^3x\,\calD \ ; & \calD &= x^\mu\calP_\mu \ ; \label{eq:D_general} \\
	&\text{Special conformal:} & K^\mu &=  \int d^3x\,\calK^\mu \ ; & \calK^\mu &= \frac{1}{2}x_\nu x^\nu \calP^\mu - x^\mu x^\nu\calP_\nu + x_\nu \calM^{\nu\mu} + \calR^\mu \ . \label{eq:K_general}
\end{align}
Here, $P^\mu$ is fixed canonically by the Lagrangian. Note that it only contains derivatives along the $\ell_\mu x^\mu = 0$ hyperplane, as the transverse derivatives cancel between the two terms in \eqref{eq:P_general}. The terms proportional to $\calP^\mu$ in \eqref{eq:J_general}-\eqref{eq:K_general} are the orbital parts of the generators. The extra $\calM^{\mu\nu}$ term in \eqref{eq:J_general} is the intrinsic part of Lorentz rotations, which includes both spin and interactions. It reappears in \eqref{eq:K_general}, expressing the local rotation of the frame under the special conformals. The $\calR^\mu$ term in \eqref{eq:K_general} is an extra intrinsic part of the special conformals. Note that the fields' length dimension is taken care of automatically, without appearing as an explicit term in \eqref{eq:D_general}-\eqref{eq:K_general}. 

The challenge of writing (e.g. cubic) interactions in the lightcone formalism is to find cubic Lagrangians $\calL_{[3]}$, \emph{together with cubic contributions to the internal generators} $\calM^{\mu\nu}_{[3]},\calR^\mu_{[3]}$, so that the algebra of the generators \eqref{eq:P_general}-\eqref{eq:K_general} closes correctly. Note that it's sufficient to check the algebra on a single hyperplane $\ell_\mu x^\mu = 0$: the generators and their algebra at other values of $\ell_\mu x^\mu$ follow automatically, by acting with the transverse component of translations $P^\mu$.

At the quadratic (free-field) order, it can be easier to express the generators \eqref{eq:P_general}-\eqref{eq:K_general} in terms of their linear action on the fields:
\begin{align}
	i[P_{[2]}^\mu,\Phi_h] &\equiv P_{\text{lin}.}^\mu\Phi_h = \left(\del^\mu - \frac{\ell^\mu\Box}{2(\ell\cdot\del)}\right)\Phi_h \ ; \label{eq:P_linear} \\
	i[J^{\mu\nu}_{[2]},\Phi_h] &\equiv J^{\mu\nu}_{\text{lin.}}\Phi_h =  \left(2x^{[\mu}P_{\text{lin.}}^{\nu]} + M^{\mu\nu}_{\text{lin.}} \right)\Phi_h \ ; \label{eq:J_linear} \\
	i[D_{[2]},\Phi_h] &\equiv D_{\text{lin.}}\Phi_h =  \left(x_\mu P_{\text{lin.}}^\mu + \Delta \right)\Phi_h \ ; \label{eq:D_linear}  \\
	i[K_{[2]}^\mu,\Phi_h] &\equiv K^\mu_{\text{lin.}}\Phi_h = \left(\frac{1}{2}x_\nu x^\nu P_{\text{lin.}}^\mu - x^\mu(x_\nu P_{\text{lin.}}^\nu + \Delta) + x_\nu M_{\text{lin.}}^{\nu\mu} + R^\mu_{\text{lin.}} \right)\Phi_h \label{eq:K_linear} \ ,
\end{align}
where we follow the same ``orbital + intrinsic'' structure as in \eqref{eq:P_general}-\eqref{eq:K_general}, but with the fields' scaling weight $\Delta$ now appearing explicitly in \eqref{eq:D_linear}-\eqref{eq:K_linear}. The intrinsic parts of the generators \eqref{eq:J_linear}-\eqref{eq:K_linear} read:
\begin{align}
	M^{\mu\nu}_{\text{lin.}} = - \frac{ih\epsilon^{\mu\nu\rho\sigma}\ell_\rho\del_\sigma}{\ell\cdot\del} \ ; \quad \Delta = 1 \ ; \quad R^\mu_{\text{lin.}} = \frac{h^2\ell^\mu}{\ell\cdot\del} \ . \label{eq:M_Delta_R_linear}
\end{align}
Note that \eqref{eq:P_linear} is again engineered so that derivatives transverse to $\ell\cdot x = \const$ cancel, while reducing to $\del^\mu$ on the free-field equation of motion $\Box\Phi_h = 0$. To see that $h$ is indeed the field's helicity, note that when $M^{\mu\nu}_{\text{lin.}}$ is contracted with a spatial bivector orthogonal to $\ell^\mu$, the derivatives in the numerator and denominator cancel, leaving just $ih$. We can now write the free-field generators themselves in a unified way, in terms of the linear transformations \eqref{eq:P_linear}-\eqref{eq:K_linear}. For each of the generators $G\in \{P^\mu,J^{\mu\nu},D,K^\mu\}$, we have simply:
\begin{align}
	G_{[2]} = \int d^3x\sum_h\Phi_{-h}(\ell\cdot\del)G_{\text{lin}.}\Phi_h \ . \label{eq:quadratic_generators}
\end{align}
This form of the generators doesn't map directly onto \eqref{eq:P_general}-\eqref{eq:K_general}, but is related through integration by parts. It's non-obvious but true that the generators \eqref{eq:quadratic_generators} indeed produce the commutators \eqref{eq:P_linear}-\eqref{eq:K_linear} (the non-obvious piece is the integration-by-parts when commuting with $\Phi_{-h}$). Note that the inverse derivatives in \eqref{eq:P_linear}-\eqref{eq:K_linear} always cancel against the $(\ell\cdot\del)$ derivative in \eqref{eq:quadratic_generators}.

It's straightforward to check that the free-field generators \eqref{eq:quadratic_generators}, or equivalently the linear transformations \eqref{eq:P_linear}-\eqref{eq:K_linear}, indeed satisfy the conformal algebra (see Appendix \ref{app:conformal}). We now turn to discuss the conformal symmetry, and its reduced Poincare/AdS versions, for cubic interactions.

\subsection{Cubic vertices in Minkowski and AdS} \label{sec:framework:cubic}

Cubic vertices of massless fields in 4d are characterized by the helicities $h_i$ ($i=1,2,3$) of the fields involved. HS gravity contains all the vertices allowed by gauge/spacetime symmetry, with the exception of the purely-scalar vertex $h_1=h_2=h_3=0$. All other vertices can be divided into two classes: chiral $h_1+h_2+h_3>0$ and anti-chiral $h_1+h_2+h_3<0$. Since these are related by complex conjugation, we will write out only the chiral sector explicitly. To specify the conformal generators in the form \eqref{eq:P_general}-\eqref{eq:K_general}, we will need the cubic pieces of the Lagrangian $\calL$, the intrinsic Lorentz rotations $\calM^{\mu\nu}$, and the special conformals $\calR^\mu$. Fortunately, an immediate simplification occurs: the cubic piece of $\calM^{\mu\nu}$ is always along the left-handed bivector $q^\alpha q^\beta$ (for chiral vertices) or the right-handed bivector $\bar q^{\dot\alpha}\bar q^{\dot\beta}$ (for anti-chiral), while the cubic piece of $\calR^\mu$ simply vanishes:
\begin{gather}
	\calM^{\alpha\beta}_{[3]} = 4q^\alpha q^\beta\calM_{[3]} \ ; \quad \calM^{\dot\alpha\dot\beta}_{[3]} = 4\bar q^{\dot\alpha}\bar q^{\dot\beta}\calM_{[3]}^\dagger \ ; \quad \calR_{[3]}^\mu = 0 \ . \label{eq:M_R_cubic}
\end{gather}
It remains to specify the cubic objects $\calL_{[3]}$ and $\calM_{[3]}$. Since they're constructed out of three fields $\Phi_{h_i}$ ($i=1,2,3$) and derivatives thereof, it is convenient to introduce spacetime derivatives $\del_\mu^{(i)}$ that act on field number $i$. It turns out that the chiral vertices depend exclusively on derivatives contracted with $q^\alpha$, i.e. derivatives along the left-handed plane $q^\alpha q^\beta$:
\begin{align}
	p^{(i)}_{\dot\alpha} \equiv q^\alpha\del^{(i)}_{\alpha\dot\alpha} \ . \label{eq:p}
\end{align}
The structure \eqref{eq:p} is tightly connected to twistor formulations of chiral theories -- see e.g. \cite{Adamo:2022lah,Mason:2025pbz,Tran:2025uad} in the higher-spin context. 

The derivatives \eqref{eq:p} can be contracted with each other, or with $\bar q^{\dot\alpha}$, which produces the lightlike derivatives $\ell\cdot\del^{(i)}$:
\begin{align}
	\bbP \equiv \frac{1}{3}\left(p^{(1)}_{\dot\alpha}p^{(2)\dot\alpha} + p^{(2)}_{\dot\alpha}p^{(3)\dot\alpha} + p^{(3)}_{\dot\alpha}p^{(1)\dot\alpha}  \right) \ ; \quad \beta_i \equiv \ell\cdot\del^{(i)} = -\bar q^{\dot\alpha} p^{(i)}_{\dot\alpha} \ . \label{eq:P_beta}
\end{align}
Note that the individual terms $p^{(i)}_{\dot\alpha}p^{(j)\dot\alpha}$ inside $\bbP$ are all related through integration by parts. Also, if desired, integration by parts can eliminate one of the $\beta_i$'s. 

\subsubsection{Minkowski vertices}

In Minkowski space, the cubic vertices take the form \cite{Bengtsson:1983pd,Fradkin:1991iy}:
\begin{align}
   \calL_{[3]} &= \sum_{h_1+h_2+h_3>0}\frac{C_{h_1h_2h_3}\calV_{h_1h_2h_3}\Phi_{h_1}\Phi_{h_2}\Phi_{h_3}}{(\ell\cdot\del^{(1)})^{h_1}(\ell\cdot\del^{(2)})^{h_2}(\ell\cdot\del^{(3)})^{h_3}} + h.c. \ ; \label{eq:L_cubic} \\
   \calM_{[3]} &= \sum_{h_1+h_2+h_3>0}\frac{C_{h_1h_2h_3}\calM_{h_1h_2h_3}\Phi_{h_1}\Phi_{h_2}\Phi_{h_3}}{(\ell\cdot\del^{(1)})^{h_1}(\ell\cdot\del^{(2)})^{h_2}(\ell\cdot\del^{(3)})^{h_3}} \ ; \label{eq:M_cubic}
\end{align}
Here, $C_{h_1h_2h_3}$ are coupling constants, and $\calV_{h_1h_2h_3},\calM_{h_1h_2h_3}$ are differential operators (with positive powers of derivatives; all the potentially negative powers are in the explicit denominators of \eqref{eq:L_cubic}-\eqref{eq:M_cubic}). In terms of the derivative combinations \eqref{eq:P_beta}, the operators $\calV_{h_1h_2h_3},\calM_{h_1h_2h_3}$ read simply:
\begin{align}
	\calV_{h_1h_2h_3} &= \bbP^{h_1+h_2+h_3} \ ; \label{eq:flat_vertex} \\ 
	\calM_{h_1h_2h_3} &= \frac{2}{3}\bbP^{h_1+h_2+h_3-1}\big((h_2-h_3)\beta_1 + (h_3-h_1)\beta_2 + (h_1-h_2)\beta_3 \big) \ . \label{eq:flat_M}
\end{align}
With these ingredients, the Poincare algebra \eqref{eq:P_general}-\eqref{eq:J_general} closes to cubic order. For $h_1+h_2+h_3>1$, the vertices \eqref{eq:flat_vertex}-\eqref{eq:flat_M} violate dilatations, since they require a dimensionful coupling constant $C_{h_1h_2h_3}$. In the special case $h_1+h_2+h_3=1$, which includes self-dual Yang-Mills and its HS generalization \cite{Ponomarev:2017nrr,Krasnov:2021nsq}, the coupling is dimensionless. In fact, as we show in Appendix \ref{app:conformal}, in this case the entire conformal group \eqref{eq:P_general}-\eqref{eq:K_general} closes to cubic order \footnote{We thank the referee for pointing out that this claim has an existing trail in the literature \cite{MetsaevThesis,Ponomarev:2017nrr}.}.

When any ``true higher-spin'' vertex is switched on, i.e. if there's any non-vanishing coupling with $h_1+h_2+h_3>2$ (and with one of the $h_i$'s negative, i.e. in the covariant language the coupling can't be constructed as just a product of linearized curvature tensors), then consistency at the quartic order \cite{Ponomarev:2016lrm,Serrani:2025owx} requires that \emph{all} chiral vertices (at least, all those with even spins) are included. Their couplings are then fixed as:
\begin{align}
	C_{h_1h_2h_3} \sim \frac{a^{h_1+h_2+h_3}}{(h_1+h_2+h_3-1)!} \ , \label{eq:c_proportionality}
\end{align}
where $a$ is a constant with units of length.

\subsubsection{AdS vertices} \label{sec:framework:cubic:AdS}

We now turn to the AdS case, as described in Metsaev's technical masterpiece \cite{Metsaev:2018xip}. To get the AdS algebra instead of the Poincare algebra, we absorb the length dimensions of the couplings $C_{h_1h_2h_3}$ into powers of the preferred spatial coordinate $z$ (for now, as in \cite{Metsaev:2018xip}, we set $\ell^\mu$ and $z$ orthogonal). Specifically, we replace \eqref{eq:L_cubic}-\eqref{eq:M_cubic} with:
\begin{align}
	\calL_{[3]} &= \sum_{h_1+h_2+h_3>0}\frac{C_{h_1h_2h_3}\calV_{h_1h_2h_3}z^{h_1+h_2+h_3-1}\Phi_{h_1}\Phi_{h_2}\Phi_{h_3}}{(\ell\cdot\del^{(1)})^{h_1}(\ell\cdot\del^{(2)})^{h_2}(\ell\cdot\del^{(3)})^{h_3}} + h.c. \ ; \label{eq:L_cubic_AdS} \\
    \calM_{[3]} &= \sum_{h_1+h_2+h_3>0}\frac{C_{h_1h_2h_3}\calM_{h_1h_2h_3}z^{h_1+h_2+h_3}\Phi_{h_1}\Phi_{h_2}\Phi_{h_3}}{(\ell\cdot\del^{(1)})^{h_1}(\ell\cdot\del^{(2)})^{h_2}(\ell\cdot\del^{(3)})^{h_3}} \ , \label{eq:M_cubic_AdS}
\end{align}
where the differential operators $\calV_{h_1h_2h_3},\calM_{h_1h_2h_3}$ now also contain derivatives of the explicit function of $z$ in \eqref{eq:L_cubic_AdS}-\eqref{eq:M_cubic_AdS}. The inclusion of the $z$ factors restores dilatations, along with the special conformals, at the cost of breaking the components of $P^\mu,J^{\mu\nu},K^\mu$ along the $z$ direction. We are thus left with the conformal group of the 3d spacetime orthogonal to $z$, namely the AdS group. Note that the statement here is rather subtle. The conformal generators \eqref{eq:P_general}-\eqref{eq:K_general} along the $z$ axis are still defined, with their cubic pieces following the structure \eqref{eq:M_R_cubic}, but they are not symmetries. In particular, there's a component of $J^{\mu\nu}$ that isn't a symmetry, but whose non-trivial cubic piece enters via \eqref{eq:K_general} into the expression for a $K^\mu$ component that \emph{is} a symmetry (in the notation of \cite{Metsaev:2018xip}, these components are $J^{-z}$ and $K^-$).

In AdS, the differential operators $\calV_{h_1h_2h_3},\calM_{h_1h_2h_3}$ are not as simple as their flat counterparts \eqref{eq:flat_vertex}-\eqref{eq:flat_M}. They are given by some polynomials of degree $h_1+h_2+h_3$ in the derivative combinations $\bbP$ and $\beta_i\del_z$, where the $\del_z$ derivative acts on the explicit function of $z$ in \eqref{eq:L_cubic_AdS}-\eqref{eq:M_cubic_AdS}:
\begin{align}
	\calV_{h_1h_2h_3} = \calV_{h_1h_2h_3}(\bbP,\beta_1\del_z,\beta_2\del_z,\beta_3\del_z) \ ; \quad \calM_{h_1h_2h_3} = \calM_{h_1h_2h_3}(\bbP,\beta_1\del_z,\beta_2\del_z,\beta_3\del_z) \ . \label{eq:AdS_vertex}
\end{align}
The flat limit is captured by the leading terms at large $z$, i.e. at small $\del_z$. In this limit, the flat vertices \eqref{eq:flat_vertex}-\eqref{eq:flat_M} are recovered, up to total factors of $z^{h_1+h_2+h_3-1}$. Thus, the leading terms of the polynomials \eqref{eq:AdS_vertex} read:
\begin{align}
	\calV_{h_1h_2h_3} &= \bbP^{h_1+h_2+h_3} + \dots \ ; \label{eq:Q_V_limit} \\
	\calM_{h_1h_2h_3} &= \frac{2\bbP^{h_1+h_2+h_3-1}\big((h_2-h_3)\beta_1 + (h_3-h_1)\beta_2 + (h_1-h_2)\beta_3\big)\del_z}{3(h_1+h_2+h_3)} + \dots \ . \label{eq:Q_M_limit}
\end{align}
Note that the appearance of $\del_z$ in the flat limit \eqref{eq:Q_M_limit} is an artifact of our choice \eqref{eq:AdS_vertex} to always group $\del_z$'s together with $\beta_i$'s, which will prove convenient below.

As in the flat case, consistency at the quartic order imposes a proportionality \eqref{eq:c_proportionality} between all the chiral couplings. However, the constant $a$ in \eqref{eq:c_proportionality} is now dimensionless (or, equivalently, is measured in units of the AdS radius). In the theory's chiral version, $a$ can always be cancelled by the asymmetric rescaling $\Phi_h\to a^{-h}\Phi_h$. However, in the real/unitary theory, where $\Phi_{\pm h}$ are related by a reality condition $\Phi_{-h} = \Phi^\dagger_h$, such rescalings are limited to complex phases $\Phi_h\to e^{ih\theta}\Phi_h$. These can reduce $a$ to a \emph{real positive} constant, but cannot get rid of it entirely. On the other hand, for type-A HS gravity, it's easy to fix the value of $a$ holographically. We perform this calculation in Appendix \ref{app:matching}, with the simple result $a = 1$, i.e.:
\begin{align}
	C_{h_1h_2h_3} \sim \frac{1}{(h_1+h_2+h_3-1)!} \ . \label{eq:a}
\end{align}

The detailed form of the polynomials $\calV_{h_1h_2h_3},\calM_{h_1h_2h_3}$ will not concern us here. It is given indirectly in \cite{Metsaev:2018xip}, via a differential equation (and a semi-explicit solution in terms of an exponentiated differential operator). Instead, we note here another useful rewriting step. Thanks to the spinor form \eqref{eq:q_bar} of the condition $\ell^\mu\del_\mu z = 0$, we can recast the $\del_z$ derivatives in a form analogous to \eqref{eq:p}:
\begin{align} 
	p^{(0)}_{\dot\alpha} \equiv q^\alpha\del^{(0)}_{\alpha\dot\alpha} = \bar q_{\dot\alpha}\del_z \ ,
\end{align}
where $\del^{(0)}_\mu$ is a derivative acting on the explicit function of $z$, in analogy with $\del^{(i)}_\mu$. The combinations $\beta_i\del_z$ that enter the vertices \eqref{eq:AdS_vertex} can now be expressed as:
\begin{align}
  \beta_i\del_z = p^{(0)}_{\dot\alpha}p^{(i)\dot\alpha} \ .
\end{align}
This means that \emph{all} the derivative combinations that enter the vertices \eqref{eq:AdS_vertex} are composed from contractions of the basic derivatives $p^{(I)}_{\dot\alpha}$, with $I=0,1,2,3$. We can thus recast \eqref{eq:AdS_vertex} as polynomials of doubled degree $2(h_1+h_2+h_3)$ in these basic derivatives:
\begin{align}
	\calV_{h_1h_2h_3} = \calV_{h_1h_2h_3}(p^{(0)}_{\dot\alpha},p^{(1)}_{\dot\alpha},p^{(2)}_{\dot\alpha},p^{(3)}_{\dot\alpha}) \ ; \quad 
	\calM_{h_1h_2h_3} = \calM_{h_1h_2h_3}(p^{(0)}_{\dot\alpha},p^{(1)}_{\dot\alpha},p^{(2)}_{\dot\alpha},p^{(3)}_{\dot\alpha}) \ , \label{eq:AdS_vertex_p}
\end{align}
with the understanding that the right-handed spinor indices are always contracted, i.e. that $\calV_{h_1h_2h_3},\calM_{h_1h_2h_3}$ are invariant under right-handed Lorentz rotations. 

\section{Generalized lightcone frames and extension to de Sitter space} \label{sec:extend}

In this section, we perform the generalization to AdS lightcone frames with $\ell^\mu\del_\mu z \neq 0$, and from there to de Sitter space. Some of the groundwork was already done in the previous section, where we rewrote Metsaev's formalism in more covariant-looking notation. The next step is to focus on the sector of chiral vertices, and shift to a field frame that's better adapted to it.

\subsection{Chiral field frame} \label{sec:extend:chiral}

Let us temporarily retreat to the chiral version of HS theory. This is the theory obtained by simply throwing away the anti-chiral vertices, i.e. the ``h.c.'' term in \eqref{eq:L_cubic} and the right-handed $\calM_{[3]}^{\dot\alpha\dot\beta}$. Since the Lagrangian is no longer real, the opposite-helicity fields $\Phi_{\pm h}$ are no longer related by complex conjugation. It is then useful to shift to an alternative field frame, in which the fields are tilted by positive/negative powers of $\ell\cdot\del$ according to their helicity:
\begin{align}
	\phi_h \equiv (\ell\cdot\del)^{-h}\Phi_h \ . \label{eq:phi}
\end{align}
Under this transformation, the kinetic Lagrangian \eqref{eq:action} and canonical commutators \eqref{eq:commutators} simply acquire a sign factor:
\begin{align}
	\calL_{[2]} &= \frac{(-1)^h}{2}\sum_h \phi_{-h}\Box\phi_h\ ;
	&[\phi_h(x),\phi_{h'}(x')] = \frac{(-1)^h\delta_{-h,h'}}{2i(\ell\cdot\del)}\,\delta^3(x - x') \ . \label{eq:commutators_chiral}
\end{align}
The quadratic (i.e. free-field) part of the generators \eqref{eq:quadratic_generators} becomes:
\begin{align}
	G_{[2]} = \int d^3x\sum_h(-1)^h\phi_{-h}(\ell\cdot\del)\tilde G_{\text{lin}.}\phi_h \ . \label{eq:quadratic_generators_chiral}
\end{align}
Here, the new linear operators $\tilde G_{\text{lin}.}$ are obtained by conjugating \eqref{eq:J_linear}-\eqref{eq:K_linear} by the transformation \eqref{eq:phi} (which leaves $P^\mu_{\text{lin.}}$ unchanged). It turns out that these new operators are simpler -- the intrinsic part of $\tilde J^{\mu\nu}_{\text{lin}.}$ is purely left-handed, and the intrinsic part of $\tilde K_{\text{lin}.}^\mu$ vanishes:
\begin{align}
	i[P_{[2]}^{\alpha\dot\alpha},\phi_h] &= P_{\text{lin}.}^{\alpha\dot\alpha}\phi_h = \left(\del^{\alpha\dot\alpha} + \frac{q^\alpha\bar q^{\dot\alpha}\Box}{q^\gamma\bar q^{\dot\gamma}\del_{\gamma\dot\gamma}}\right)\phi_h \ ; \label{eq:P_linear_chiral} \\
	i[J^{\alpha\beta}_{[2]},\phi_h] &= \tilde J^{\alpha\beta}_{\text{lin.}}\phi_h =  \left(-x^{(\alpha}{}_{\dot\alpha}P_{\text{lin.}}^{\beta)\dot\alpha} + \tilde M^{\alpha\beta}_{\text{lin.}} \right)\phi_h \ ; \label{eq:J_left_linear_chiral} \\
	i[J^{\dot\alpha\dot\beta}_{[2]},\phi_h] &= \tilde J^{\dot\alpha\dot\beta}_{\text{lin.}}\phi_h =  -x_\alpha{}^{(\dot\alpha}P_{\text{lin.}}^{|\alpha|\dot\beta)} \phi_h \ ; \label{eq:J_right_linear_chiral} \\
	i[D_{[2]},\phi_h] &= \tilde D_{\text{lin.}}\phi_h =  \left(-\frac{1}{2}x_{\alpha\dot\alpha} P_{\text{lin.}}^{\alpha\dot\alpha} + \tilde\Delta \right)\phi_h \ ; \label{eq:D_linear_chiral}  \\
	i[K_{[2]}^{\alpha\dot\alpha},\phi_h] &= \tilde K^{\alpha\dot\alpha}_{\text{lin.}}\phi_h = \left(-\frac{1}{4}x_{\beta\dot\beta} x^{\beta\dot\beta} P_{\text{lin.}}^{\alpha\dot\alpha} 
	  - x^{\alpha\dot\alpha}\left(- \frac{1}{2}x_{\beta\dot\beta} P_{\text{lin.}}^{\beta\dot\beta} + \tilde\Delta\right) - \frac{1}{2}x_\beta{}^{\dot\alpha}\tilde M_{\text{lin.}}^{\beta\alpha} \right)\phi^h \label{eq:K_linear_chiral} \ ,
\end{align}
where the intrinsic Lorentz and scaling weights are given by:
\begin{align}
	\tilde M^{\alpha\beta}_{\text{lin.}} = \frac{4h\bar q^{\dot\alpha} q^{(\alpha}\del^{\beta)}{}_{\dot\alpha}}{q^\gamma\bar q^{\dot\gamma}\del_{\gamma\dot\gamma}} \ ; \quad  \tilde\Delta = 1-h \ . \label{eq:M_Delta_chiral}
\end{align}
Let us now turn to the cubic vertices. Since we keep only the chiral ones, eq. \eqref{eq:M_R_cubic} becomes:
\begin{align}
  \calM_{[3]}^{\alpha\beta} = 4q^\alpha q^\beta\calM_{[3]} \ ; \quad \calM_{[3]}^{\dot\alpha\dot\beta} = \calR^\mu_{[3]} = 0 \ . \label{eq:M_R_chiral}
\end{align}
In the new field frame \eqref{eq:phi}, the lightlike derivatives in the denominator of \eqref{eq:L_cubic}-\eqref{eq:M_cubic} cancel, so the vertices $\calL_{[3]},\calM_{[3]}$ become simply:
\begin{align}
	\calL_{[3]} &= \sum_{h_1+h_2+h_3>0}C_{h_1h_2h_3}\calV_{h_1h_2h_3}(p^{(0)}_{\dot\alpha},p^{(1)}_{\dot\alpha},p^{(2)}_{\dot\alpha},p^{(3)}_{\dot\alpha})z^{h_1+h_2+h_3-1}\phi_{h_1}\phi_{h_2}\phi_{h_3} \ ; \label{eq:L_cubic_chiral} \\
	\calM_{[3]} &= \sum_{h_1+h_2+h_3>0}C_{h_1h_2h_3}\calM_{h_1h_2h_3}(p^{(0)}_{\dot\alpha},p^{(1)}_{\dot\alpha},p^{(2)}_{\dot\alpha},p^{(3)}_{\dot\alpha})z^{h_1+h_2+h_3}\phi_{h_1}\phi_{h_2}\phi_{h_3} \ , \label{eq:M_cubic_chiral}
\end{align}
where the differential operators $\calV_{h_1h_2h_3},\calM_{h_1h_2h_3}$ are the same as in section \ref{sec:framework:cubic:AdS}. 

Now, let us plug \eqref{eq:M_R_chiral} into the generators \eqref{eq:P_general}-\eqref{eq:K_general}. Their cubic part (in spinor indices) then reads:
\begin{align}
	\calP_{[3]}^{\alpha\dot\alpha} &= -2q^\alpha\bar q^{\dot\alpha}\calL_{[3]} \ ; \label{eq:P_cubic} \\
	\calJ_{[3]}^{\alpha\beta} &=  -2q^{(\alpha}\bar q_{\dot\alpha}x^{\beta)\dot\alpha} \calL_{[3]} + 4q^\alpha q^\beta\calM_{[3]} \ ; \label{eq:J_left_cubic} \\
	\calJ_{[3]}^{\dot\alpha\dot\beta} &= 2q^\alpha\bar q^{(\dot\alpha}x_\alpha{}^{\dot\beta)} \calL_{[3]} \ ; \label{eq:J_right_cubic} \\
	\calD_{[3]} &=  q_\alpha\bar q_{\dot\alpha}x^{\alpha\dot\alpha}\calL_{[3]} \ ; \label{eq:D_cubic} \\
	\calK_{[3]}^{\alpha\dot\alpha} &= \left(\frac{1}{2}q^\alpha q^{\dot\alpha}x_{\beta\dot\beta}x^{\beta\dot\beta} - q_\beta q_{\dot\beta}x^{\beta\dot\beta} x^{\alpha\dot\alpha} \right)\calL_{[3]} 
    	+ 2q^\alpha q_\beta x^{\beta\dot\alpha} \calM_{[3]} \ . \label{eq:K_cubic}
\end{align}
We can now use the vertices' structure \eqref{eq:L_cubic_chiral}-\eqref{eq:M_cubic_chiral} to work out the action of the generators \eqref{eq:P_cubic}-\eqref{eq:K_cubic} on the fields $\phi_h$. For any generator $G = \int d^3x\,\calG$, its action on the fields can be expressed as a variational derivative, via the canonical commutation relations \eqref{eq:commutators_chiral}:
\begin{align}
	i[G,\phi_h] = \frac{(-1)^h}{2\ell\cdot\del}\frac{\delta\calG}{\delta\phi_{-h}} \ . \label{eq:variational}
\end{align}
Applying this to \eqref{eq:P_cubic}-\eqref{eq:K_cubic}, we need to carefully consider the effects of derivatives on the explicit $x^\mu$ factors in \eqref{eq:J_left_cubic}-\eqref{eq:K_cubic}. The inverse $\ell\cdot\del$ derivative in \eqref{eq:variational} commutes with all these factors, but the derivatives arising from $\frac{\delta}{\delta\phi_{-h}}$ do not. However, the effect of such derivatives on the $x^\mu$ factors ends up rather simple, thanks to the fact that $\calL_{[3]},\calM_{[3]}$ contain derivatives only in the form $p_{\dot\alpha} = q^\alpha\del_{\alpha\dot\alpha}$: this leads to the vanishing of some contributions,via contractions of $q^\alpha$ with itself. The non-vanishing contributions all involve just \emph{one} derivative acting on an $x^\mu$ factor, and come from the $\calL_{[3]}$ terms in \eqref{eq:J_left_cubic} and \eqref{eq:K_cubic}. They can be nicely packaged by incorporating a (helicity-dependent) piece of $\calL_{[3]}$ into $\calM_{[3]}$:
\begin{align}
 \begin{split}
	\calM_{[3]}^{(h)} \equiv{}& \calM_{[3]} - \bar q_{\dot\alpha}\sum_{h_1+h_2+h_3>0}C_{h_1h_2h_3}
	  \left(\delta_{hh_1}\frac{\del}{\del p^{(1)}_{\dot\alpha}} + \delta_{hh_2}\frac{\del}{\del p^{(2)}_{\dot\alpha}} + \delta_{hh_3}\frac{\del}{\del p^{(3)}_{\dot\alpha}}\right) \\
	 &\qquad\qquad\qquad \times \calV_{h_1h_2h_3}(p^{(0)},p^{(1)},p^{(2)},p^{(3)}) z^{h_1+h_2+h_3-1}\phi_{h_1}\phi_{h_2}\phi_{h_3} \ .
 \end{split}
\end{align}
The action of \eqref{eq:P_cubic}-\eqref{eq:K_cubic} on the fields can then be expressed as:
\begin{align}
	i[P_{[3]}^{\alpha\dot\alpha},\phi_h] &= \frac{(-1)^h q^\alpha\bar q^{\dot\alpha}}{q^\gamma\bar q^{\dot\gamma}\del_{\gamma\dot\gamma}}\frac{\delta\calL_{[3]}}{\delta\phi_{-h}} \ ; \label{eq:P_cubic_action} \\
	i[J_{[3]}^{\alpha\beta},\phi_h] &=  \frac{(-1)^h}{q^\gamma\bar q^{\dot\gamma}\del_{\gamma\dot\gamma}}\left(q^{(\alpha}\bar q_{\dot\alpha}x^{\beta)\dot\alpha} \frac{\delta\calL_{[3]}}{\delta\phi_{-h}} 
	   - 2q^\alpha q^\beta\frac{\delta\calM^{(-h)}_{[3]}}{\delta\phi_{-h}} \right) \ ; \label{eq:J_left_cubic_action} \\
	i[J_{[3]}^{\dot\alpha\dot\beta},\phi_h] &= \frac{(-1)^{h+1} q^\alpha\bar q^{(\dot\alpha}x_\alpha{}^{\dot\beta)}}{q^\gamma\bar q^{\dot\gamma}\del_{\gamma\dot\gamma}} \frac{\delta\calL_{[3]}}{\delta\phi_{-h}} \ ; \label{eq:J_right_cubic_action} \\
	i[D_{[3]},\phi_h] &=  \frac{(-1)^{h+1} q_\alpha\bar q_{\dot\alpha}x^{\alpha\dot\alpha}}{2q^\gamma\bar q^{\dot\gamma}\del_{\gamma\dot\gamma}}\frac{\delta\calL_{[3]}}{\delta\phi_{-h}} \ ; \label{eq:D_cubic_action} \\
	i[K_{[3]}^{\alpha\dot\alpha},\phi_h] &= \frac{(-1)^{h+1}}{2q^\gamma\bar q^{\dot\gamma}\del_{\gamma\dot\gamma}}\left(
	  \left(\frac{1}{2}q^\alpha\bar q^{\dot\alpha}x_{\beta\dot\beta}x^{\beta\dot\beta} - q_\beta\bar q_{\dot\beta}x^{\beta\dot\beta} x^{\alpha\dot\alpha} \right) \frac{\delta\calL_{[3]}}{\delta\phi_{-h}}
	    + 2q^\alpha q_\beta x^{\beta\dot\alpha} \frac{\delta\calM^{(-h)}_{[3]}}{\delta\phi_{-h}} \right) \ . \label{eq:K_cubic_action}
\end{align}

\subsection{Summary of the chiral theory: action and symmetry transformations} \label{sec:extend:summary}

Let us now summarize our knowledge of the chiral AdS theory, in a way that will facilitate the extension to $\ell^\mu\del_\mu z\neq 0$ in the next subsection \ref{sec:extend:analytic_continuation}. 

The chiral vertices are encoded in the differential operators  $\calV_{h_1h_2h_3},\calM_{h_1h_2h_3}$; these are polynomials w.r.t. the basic derivative operators $p^{(I)}_{\dot\alpha} = q^\alpha\del_{\alpha\dot\alpha}^{(I)}$ that act either on one of the three fields (for $I=1,2,3$) or on the explicit function of $z$ (for $I=0$). The action to cubic order reads:
\begin{align}
	S &= \int d^4x\left(\frac{(-1)^h}{2}\sum_h \phi_{-h}\Box\phi_h + \calL_{[3]}\right) \ ; \label{eq:action_chiral} \\
	\calL_{[3]} &= \sum_{h_1+h_2+h_3>0}C_{h_1h_2h_3}\calV_{h_1h_2h_3}z^{h_1+h_2+h_3-1}\phi_{h_1}\phi_{h_2}\phi_{h_3} \ , \label{eq:L_cubic_chiral_again}
\end{align}
with field equations:
\begin{align}
	\Box\phi_h = (-1)^{h+1}\frac{\delta\calL_{[3]}}{\delta\phi_{-h}} \ . \label{eq:field_eq}
\end{align}
With the field equations imposed, we can write the effect of translations $P^\mu$ (including the component transverse to the lightlike hyperplane) as simply $\del^\mu$. Overall, with the field equations imposed, the effect of the symmetry generators \eqref{eq:P_linear_chiral}-\eqref{eq:K_linear_chiral},\eqref{eq:P_cubic_action}-\eqref{eq:K_cubic_action} on the fields $\phi_h$ can be packaged as:
\begin{align}
   i[P^{\alpha\dot\alpha},\phi_h] &= \del^{\alpha\dot\alpha}\phi_h \ ; \label{eq:P_action} \\
   i[J^{\alpha\beta},\phi_h] &= -x^{(\alpha}{}_{\dot\alpha}\del^{\beta)\dot\alpha}\phi_h + \tilde M^{\alpha\beta} \ ; \label{eq:J_left_action} \\
   i[J^{\dot\alpha\dot\beta},\phi_h] &= x^{\alpha(\dot\alpha}\del_\alpha{}^{\dot\beta)} \phi_h \ ; \label{eq:J_right_action} \\
   i[D,\phi_h] &= \left(-\frac{1}{2}x^{\alpha\dot\alpha} \del_{\alpha\dot\alpha} + \tilde\Delta \right)\phi_h \ ; \label{eq:D_action}  \\
   i[K^{\alpha\dot\alpha},\phi_h] &= \left(-\frac{1}{4}x_{\beta\dot\beta} x^{\beta\dot\beta} \del^{\alpha\dot\alpha} 
   - x^{\alpha\dot\alpha}\left(- \frac{1}{2}x^{\beta\dot\beta} \del_{\beta\dot\beta} + \tilde\Delta\right)\right)\phi_h - \frac{1}{2}x_\beta{}^{\dot\alpha}\tilde M^{\beta\alpha} \label{eq:K_action} \ ,
\end{align}
where the fields' length dimension is $\tilde\Delta = 1-h$, and the intrinsic left-handed Lorentz transformation $\tilde M^{\alpha\beta}$ is given by a sum of free+interacting pieces:
\begin{align}
	\tilde M^{\alpha\beta} = \frac{4}{(q^\delta\del_\delta{}^{\dot\gamma} z)q^\gamma \del_{\gamma\dot\gamma}}\left(h(q^\epsilon\del_{\epsilon}{}^{\dot\alpha}z) q^{(\alpha}\del^{\beta)}{}_{\dot\alpha}\phi_h - \frac{(-1)^h}{2}q^\alpha q^\beta\frac{\delta\calM^{(-h)}_{[3]}}{\delta\phi_{-h}} \right) \ ,
	\label{eq:tilde_M}
\end{align}
where:
\begin{align}
	\calM_{[3]}^{(h)} ={}& \sum_{h_1+h_2+h_3>0}C_{h_1h_2h_3}\left( \vphantom{\frac{\del}{\del p^{(1)}_{\dot\alpha}}} \calM_{h_1h_2h_3}z^{h_1+h_2+h_3}\phi_{h_1}\phi_{h_2}\phi_{h_3} \right. \label{eq:M_3_h} \\
	  &\left.\quad {} - (q^\alpha\del_{\alpha\dot\alpha}z)\left(\delta_{hh_1}\frac{\del}{\del p^{(1)}_{\dot\alpha}} + \delta_{hh_2}\frac{\del}{\del p^{(2)}_{\dot\alpha}} + \delta_{hh_3}\frac{\del}{\del p^{(3)}_{\dot\alpha}}\right)
	    \calV_{h_1h_2h_3} z^{h_1+h_2+h_3-1}\phi_{h_1}\phi_{h_2}\phi_{h_3} \right) \ . \nonumber
\end{align}
Note that we switched entirely to spinor indices, so $\ell^\mu$ appears only through its spinor constituents $q^\alpha,\bar q^{\dot\alpha}$. Moreover, we used \eqref{eq:q_bar} to replace all instances of $\bar q^{\dot\alpha}$ with $q^\alpha\del_\alpha{}^{\dot\alpha}z$ (there were only three such instances -- two in \eqref{eq:tilde_M} and one in \eqref{eq:M_3_h}). Thus, the only Lorentz-violating objects in our formulation are $q^\alpha$ and $z$. AdS symmetry in this formulation consists of two statements:
\begin{enumerate}
	\item The $z$-orthogonal components of the field transformations \eqref{eq:P_action}-\eqref{eq:K_action} preserve the action \eqref{eq:action_chiral} and field equations \eqref{eq:field_eq}.
	\item Under the field equations \eqref{eq:field_eq}, the $z$-orthogonal components of \eqref{eq:P_action}-\eqref{eq:K_action} satisfy the correct commutators of the 4d conformal algebra, thus forming its AdS subalgebra.
\end{enumerate}
Note that this framing of the chiral theory is ``almost'' manifestly local: the only inverse derivative that appears is the prefactor in \eqref{eq:tilde_M}. 

\subsection{Analytic continuation} \label{sec:extend:analytic_continuation}

Let's now describe how the chiral AdS theory of section \ref{sec:extend:summary} can be extended away from the restriction $\ell^\mu\del_\mu z = 0$. In over-simplified form, the idea is: 
\begin{enumerate}
	\item Recall that $\ell^\mu\del_\mu z = 0$ is equivalent to $\bar q^{\dot\alpha} = q^\alpha\del_\alpha{}^{\dot\alpha}z$. 
	\item Replace $\bar q^{\dot\alpha}$ with $q^\alpha\del_\alpha{}^{\dot\alpha}z$ everywhere, as we've done in section \ref{sec:extend:summary}. 
	\item With $\bar q^{\dot\alpha}$ eliminated from the formulation, the AdS symmetry is satisfied regardless of whether or not it's equal to $q^\alpha\del_\alpha{}^{\dot\alpha}z$, i.e. regardless of the $\ell^\mu\del_\mu z = 0$ restriction!
\end{enumerate}
Let us now run the argument more properly. In spinor language, the restriction $\ell^\mu\del_\mu z = 0$ becomes a \emph{reality condition}: the condition that $q^\alpha\del_\alpha{}^{\dot\alpha}z$ is the complex conjugate of $q^\alpha$, or, equivalently, that the null vector $q^\alpha q^\beta\del_\beta{}^{\dot\alpha}z$ is real. The invariance of the action \eqref{eq:action_chiral} under the ($z$-orthogonal part of) the transformations \eqref{eq:P_action}-\eqref{eq:K_action} is a complex-analytic statement, and thus does not depend on such reality conditions. The same is true of the fact that the on-shell commutators of \eqref{eq:P_action}-\eqref{eq:K_action} form the AdS algebra. 

Now, let us choose a lightlike vector $\ell^\mu = -\sigma^\mu_{\alpha\dot\alpha}q^\alpha\bar q^{\dot\alpha}$ that isn't orthogonal to $z$. We then have $\bar q^{\dot\alpha}$ not along $q^\alpha\del_\alpha{}^{\dot\alpha}z$, so that $q^\alpha q^\beta\del_\beta{}^{\dot\alpha}z$ is a \emph{complex} null vector, distinct from the real null vector $\ell^\mu$. Now, consider the chiral theory as defined in section \ref{sec:extend:summary}. As argued above, it is still symmetric under the AdS algebra. But can it still be considered a lightcone formulation w.r.t. the real $\ell^\mu$? We argue that the answer is yes, with some work.

For the action \eqref{eq:action_chiral}, no work is required. Indeed, the kinetic term is trivially the same regardless of $\ell^\mu$. As for the interaction term \eqref{eq:L_cubic_chiral_again}, it only contains derivatives of the form $q^\alpha\del_{\alpha\dot\alpha}$, which are still along the $\ell_\mu x^\mu = \const$ hyperplanes, since we still have $\ell^{\alpha\dot\alpha}\sim q^\alpha$. Thus, the action is that of a lightcone theory along $\ell^\mu$. Let us now turn to the symmetry transformations \eqref{eq:P_action}-\eqref{eq:tilde_M}. Here, we run into a problem: when $q^\alpha\del_\alpha{}^{\dot\alpha}z$ is not along $\bar q^{\dot\alpha}$, the free-field (i.e. first) term in the intrinsic Lorentz transformation \eqref{eq:tilde_M} differs from the expected one \eqref{eq:M_Delta_chiral} for a lightcone theory along $\ell^\mu$. However, we can use the Fierz identity to rearrange:
\begin{align}
	\frac{(q^\delta\del_\delta{}^{\dot\alpha}z)\del^\alpha{}_{\dot\alpha}}{(q^\gamma\del_\gamma{}^{\dot\beta} z)q^\beta\del_{\beta\dot\beta}} = \frac{\bar q^{\dot\alpha}\del^\alpha{}_{\dot\alpha}}{\bar q^{\dot\beta}q^\beta\del_{\beta\dot\beta}}
	  + \frac{(q^\delta\bar q^{\dot\delta}\del_{\delta\dot\delta}z) q^\alpha\Box}{((q^\gamma\del_{\gamma}{}^{\dot\beta} z)q^\beta\del_{\beta\dot\beta})(\bar q^{\dot\gamma}q^\gamma \del_{\gamma\dot\gamma})} \ .
\end{align}
Using this and the field equation \eqref{eq:field_eq}, the intrinsic Lorentz transformation \eqref{eq:tilde_M} becomes:
\begin{align}
	\tilde M^{\alpha\beta} = \frac{4h\bar q^{\dot\alpha} q^{(\alpha}\del^{\beta)}{}_{\dot\alpha}}{q^\gamma\bar q^{\dot\gamma}\del_{\gamma\dot\gamma}}\phi_h 
	  - \frac{4(-1)^h q^\alpha q^\beta}{(q^\delta\del_\delta{}^{\dot\gamma} z)q^\gamma \del_{\gamma\dot\gamma}}
	        \left(\frac{h(q^\epsilon\bar q^{\dot\epsilon}\del_{\epsilon\dot\epsilon}z)}{q^\xi\bar q^{\dot\xi}\del_{\xi\dot\xi}}\frac{\delta\calL_{[3]}}{\delta\phi_{-h}} + \frac{1}{2}\frac{\delta\calM^{(-h)}_{[3]}}{\delta\phi_{-h}} \right) \ ,
	\label{eq:tilde_M_continued}
\end{align}
where the free-field (first) term now agrees with \eqref{eq:M_Delta_chiral}. With this rewriting of \eqref{eq:tilde_M}, our generalization of the lightcone chiral theory to $\ell^\mu\del_\mu z \neq 0$ is complete. 

Note that, as in the original lightcone formalism, the non-linear term in the intrinsic Lorentz generator \eqref{eq:tilde_M_continued} is still along $q^\alpha q^\beta$. In fact, the generalized formalism contains just one truly non-standard feature: the non-linear term in \eqref{eq:tilde_M_continued} contains inverse derivatives not only along the real lightlike $\ell^\mu$, but also along the null spatial $q^\alpha\del_\alpha{}^{\dot\alpha}z$. This brings into question the locality and/or causality of our newly generalized lightcone formalism. In section \ref{sec:geometry}, we will show an alternative way to switch between lightcone frames, which avoids the offending inverse derivatives, and is manifestly local (in the chiral case). Building on this, we will address causality in section \ref{sec:causality}.

Now that the theory is generalized to $\ell^\mu\del_\mu z \neq 0$, the further analytic continuation from AdS to de Sitter is trivial: we simply replace the warp factor $z$ with $it$. The AdS symmetry, which was really just the $z$-orthogonal components of the conformal algebra, now becomes the $t$-orthogonal components, namely de Sitter symmetry. Note that $z\to it$ flips the overall sign of the curved metric, i.e. $\eta_{\mu\nu}/z^2$ becomes
$-\eta_{\mu\nu}/t^2$ rather than $+\eta_{\mu\nu}/t^2$, but this doesn't matter: the symmetry of both metrics is the same, and all our formulas are in terms of the flat $\eta_{\mu\nu}$ anyway. Since the overall power of $z$ (counting also its gradient $\del_{\alpha\dot\alpha} z$) in the vertices \eqref{eq:L_cubic_chiral_again},\eqref{eq:tilde_M_continued} is always $h_1+h_2+h_3-1$, the factors of $i$ resulting from $z\to it$ can be incorporated into the complex phases of the fields, and of the overall chiral coupling. Thus, an alternative way to obtain the de Sitter theory is to just substitute $z\to t$, along with the redefinitions:
\begin{align}
	\phi_h\to i^h\phi_h \ ; \quad C_{h_1h_2h_3}\to -iC_{h_1h_2h_3} \ . \label{eq:phase_rotations_dS}
\end{align}
This is consistent with the analytic continuation from higher-spin AdS/CFT to dS/CFT, as described in \cite{Anninos:2011ui}. Indeed, in AdS/CFT, the couplings $C_{h_1h_2h_3}$ are proportional to $1/\sqrt{N}$, where $N$ is the number of colors in the boundary vector model. Thus, the couplings' phase rotation in \eqref{eq:phase_rotations_dS} is equivalent to the ``$N\to -N$'' sign flip in \cite{Anninos:2011ui}, realized in practice as $O(N)\to Sp(N)$. Moreover, the phase rotation of $\phi_h$ in \eqref{eq:phase_rotations_dS} has the feature that opposite helicities $\phi_{\pm s}$ are rotated by the same phase iff the spin $s$ is even. Since the boundary currents $J^{(s)}$ are a combination of both helicities, this is consistent with the claim that analytic continuation from AdS/CFT to dS/CFT only goes through for the minimal (even-spin) version of the theory.

After describing the analytic continuations from $\ell^\mu\del_\mu z = 0$ to $\ell^\mu\del_\mu z \neq 0$ to de Sitter, we must emphasize that not all the structures from the previous sections survive them. In particular, the Hamiltonian structure of canonical commutators \eqref{eq:commutators},\eqref{eq:commutators_chiral} and integrals \eqref{eq:P_general}-\eqref{eq:K_general} on a $\ell_\mu x^\mu = \const$ hyperplane no longer makes sense when $\ell^\mu$ is replaced by the complex null vector $q^\alpha\del_\alpha{}^{\dot\alpha}z$. This is why we performed the analytic continuation on the specific formulation of section \ref{sec:extend:summary}, where this Hamiltonian structure is never used, and the symmetry generators are defined by their action on the fields.

\subsection{Back to the real theory} \label{sec:extend:real}

So far, we've analytically continued the chiral version of cubic HS theory. It is now straightforward to return to the real/unitary theory, i.e. to include the anti-chiral vertices:
\begin{enumerate}
  \item First, we reverse the field-frame shift \eqref{eq:phi} via $\Phi_h = (\ell\cdot\del)^h\phi_h$, using the real lightlike $\ell^\mu$. This brings the free-field action and symmetry transformations back to their real form \eqref{eq:action},\eqref{eq:P_linear}-\eqref{eq:M_Delta_R_linear}.
  \item Then, we add complex conjugates to all the cubic-vertex terms. This doesn't ruin the (A)dS symmetry: if the chiral terms satisfy the symmetry correctly, then so do their anti-chiral complex conjugates. And since we're working to leading order in the interactions, the two sectors never interact non-linearly.
\end{enumerate}
Taking the de Sitter version for concreteness, and applying these steps to eqs. \eqref{eq:action_chiral}-\eqref{eq:K_action},\eqref{eq:M_3_h},\eqref{eq:tilde_M_continued}, we can summarize the action, field equations and symmetries of the real theory as:
\begin{align}
	S &= \int d^4x\left(\frac{1}{2}\sum_h \Phi_{-h}\Box\Phi_h + \calL_{[3]}\right) \ ; \label{eq:action_real} \\
	\calL_{[3]} &= \sum_{h_1+h_2+h_3>0}C_{h_1h_2h_3}\calV_{h_1h_2h_3} \frac{t^{h_1+h_2+h_3-1}\Phi_{h_1}\Phi_{h_2}\Phi_{h_3}}{(\ell\cdot\del^{(1)})^{h_1}(\ell\cdot\del^{(2)})^{h_2}(\ell\cdot\del^{(3)})^{h_3}} + c.c. \ ; \label{eq:L_cubic_real}
\end{align}
\begin{align}
	\Box\Phi_h = -\frac{\delta\calL_{[3]}}{\delta\Phi_{-h}} \ ; \label{eq:field_eq_real}
\end{align}
\begin{align}
	i[P^\mu,\Phi_h] &= \del^\mu\Phi_h \ ; \label{eq:P_action_real} \\
	i[J^{\mu\nu},\Phi_h] &= 2x^{[\mu}\del^{\nu]}\Phi_h + M^{\mu\nu} \ ; \label{eq:J_action_real} \\
	i[D,\Phi_h] &= (x^\mu\del_\mu + 1)\Phi_h \ ; \label{eq:D_action_real} \\
	i[K^\mu,\Phi_h] &= \left(\frac{1}{2}x_\nu x^\nu \del^\mu - x^\mu(x^\nu\del_\nu + 1)\right)\Phi_h + x_\nu M^{\nu\mu} + \frac{h^2\ell^\mu}{\ell\cdot\del}\Phi_h \ ; \label{eq:K_action_real}
\end{align}
\begin{align}
	M^{\mu\nu} = - \frac{ih\epsilon^{\mu\nu\rho\sigma}\ell_\rho\del_\sigma}{\ell\cdot\del} - \left(\frac{\sigma^\mu_\alpha{}^{\dot\alpha}\sigma^\nu_{\beta\dot\alpha} q^\alpha q^\beta}{(q^\delta\del_\delta{}^{\dot\gamma} t)q^\gamma \del_{\gamma\dot\gamma}}
	\left(\frac{h(\ell^\rho\del_\rho t)}{\ell\cdot\del}\frac{\delta\calL_{[3]}}{\delta\Phi_{-h}} + \frac{1}{2}\frac{\delta\calM^{(-h)}_{[3]}}{\delta\Phi_{-h}} \right) + c.c. \right) \ ;
	\label{eq:M_real}
\end{align}
\begin{align}
	&\calM_{[3]}^{(h)} = \sum_{h_1+h_2+h_3>0}C_{h_1h_2h_3}\left( \vphantom{\frac{\del}{\del p^{(1)}_{\dot\alpha}}} \calM_{h_1h_2h_3}\frac{t^{h_1+h_2+h_3}\Phi_{h_1}\Phi_{h_2}\Phi_{h_3}}{(\ell\cdot\del^{(1)})^{h_1}(\ell\cdot\del^{(2)})^{h_2}(\ell\cdot\del^{(3)})^{h_3}} \right. \label{eq:M_3_h_real} \\
	&\left.{} - (q^\alpha\del_{\alpha\dot\alpha}t)\left(\delta_{hh_1}\frac{\del}{\del p^{(1)}_{\dot\alpha}} + \delta_{hh_2}\frac{\del}{\del p^{(2)}_{\dot\alpha}} + \delta_{hh_3}\frac{\del}{\del p^{(3)}_{\dot\alpha}}\right)
	\calV_{h_1h_2h_3} \frac{t^{h_1+h_2+h_3}\Phi_{h_1}\Phi_{h_2}\Phi_{h_3}}{(\ell\cdot\del^{(1)})^{h_1}(\ell\cdot\del^{(2)})^{h_2}(\ell\cdot\del^{(3)})^{h_3}} \right) \ , \nonumber
\end{align}
where the polynomial differential operators $\calV_{h_1h_2h_3}(p^{(I)}_{\dot\alpha}),\calM_{h_1h_2h_3}(p^{(I)}_{\dot\alpha})$ with $p^{(I)}_{\dot\alpha} = q^\alpha\del^{(I)}_{\alpha\dot\alpha}$ are the same as before.

\section{Geometric meaning of the new lightcone frames} \label{sec:geometry}

So far, we've been working in Poincare coordinates, paying little explicit attention to the curved (A)dS geometry. In this section, we look at the geometric meaning of the new de Sitter lightcone formalism. While we work with de Sitter space for concreteness, the discussion applies equally to the new AdS lightcone frames with $\ell^\mu\del_\mu z\neq 0$. 

\subsection{The ``null hyperplanes'' are lightcones of bulk points} \label{sec:geometry:lightcones}

In the lightcone formalism, spacetime is foliated into ``lightlike hyperplanes'' $\ell_\mu x^\mu = \const$. In the Minkowski case, these really are parallel flat hyperplanes, which can be viewed as the \emph{lightcones of points along a lightray at lightlike infinity $\scri$}. In Metsaev's AdS formalism with $\ell^\mu\del_\mu z = 0$, we no longer have flat hyperplanes, but the hypersurfaces $\ell_\mu x^\mu = \const$ are still the lightcones of points along a lightray at the conformal boundary. To see this clearly, consider the conformal boundary $z=0$. In the conformal frame induced by the Poincare coordinates, this boundary becomes a 3d Minkowski space, with its own (2d) lightlike infinity. On the boundary, the foliation leaves $\ell_\mu x^\mu = \const$ are just parallel lightlike planes, i.e. the lightcones of points on a lightray at lightlike infinity. The full $\ell_\mu x^\mu = \const$ hypersurfaces are these same lightcones, but extended into the bulk.

Now, consider the de Sitter formalism, where $z$ is replaced by $t$, necessarily with $\ell^\mu\del_\mu t \neq 0$. To clearly see the de Sitter geometry, we introduce a flat 5d spacetime $\bbR^{1,4}$, parameterized by lightcone coordinates $(u,v,\mathbf{r})$ with metric $ds^2 = -du dv + \mathbf{dr}^2$. Here, $u$ and $v$ are lightlike, and $\mathbf{r}\in\bbR^3$ is an ordinary Euclidean vector. De Sitter space $dS_4$ is then the hyperboloid $-uv+\mathbf{r}^2 = 1$ within $\bbR^{1,4}$. The Poincare coordinates $x^\mu = (t,\mathbf{x})$ can be embedded into the 5d coordinates $(u,v,\mathbf{r})$ as:
\begin{align}
	(u,v,\mathbf{r}) = -\frac{1}{t}\big(1, \mathbf{x}^2 - t^2, \mathbf{x}\big) \ . \label{eq:uv_from_Poincare}
\end{align}
Now, consider a constant lightlike vector $\ell^\mu = (1,\boldsymbol{\ell})$ in Poincare coordinates, where $\boldsymbol{\ell}\in S_2$ is some unit 3d vector. The hypersurfaces $\ell_\mu x^\mu = a$ then become:
\begin{align}
	t = \mathbf{x}\cdot\boldsymbol{\ell} - a \quad \Longrightarrow \quad (u,v,\mathbf{r}) = \frac{\big(1,2a(\mathbf{x}\cdot\boldsymbol{\ell}) - a^2 + \mathbf{x}_\perp^2, \mathbf{x}\big)}{a - \mathbf{x}\cdot\boldsymbol{\ell}} \ ,
\end{align}
where $\mathbf{x}_\perp \equiv \mathbf{x} - (\mathbf{x}\cdot\boldsymbol{\ell})\boldsymbol{\ell}$ are the components of $\mathbf{x}$ orthogonal to $\boldsymbol{\ell}$. Now, the asymptotic origin point of the $\ell_\mu x^\mu = a$ ``hyperplane'' is at $\mathbf{x}\cdot\boldsymbol{\ell}\to-\infty$, with $\mathbf{x}_\perp$ held finite. In the 5d coordinates, this becomes:
\begin{align}
	(u,v,\mathbf{r}) \to (0,-2a,-\boldsymbol{\ell}) \ . \label{eq:hyperplane_origin}
\end{align}
Thus, in the curved de Sitter geometry, the $\ell_\mu x^\mu = a$ ``hyperplane'' is just the lightcone of the completely ordinary bulk spacetime point \eqref{eq:hyperplane_origin}. Moreover, when we vary $a$ to obtain a foliation into ``parallel hyperplanes'' $\ell_\mu x^\mu = a$, the lightcones' origin points \eqref{eq:hyperplane_origin} sweep out the lightray $(u=0,\mathbf{r}=-\boldsymbol{\ell})$. Thus, the geometric meaning of our generalized lightcone frames is that, as before, they foliate spacetime into the lightcones of points that lie on a lightray, but this lightray is now in the bulk, not on the boundary.

\subsection{Locally transforming between the new lightcone frames} \label{sec:geometry:transforming}

We now present a perspective on transformations between lightcone frames, which builds on an idea from \cite{Neiman:2024vit}. Using the geometry of section \ref{sec:geometry:lightcones}, we can think of a lightcone frame in our generalized formalism as three pieces of information:
\begin{itemize}
	\item A bulk point in $dS_4$ whose lightcone serves as our ``initial'' hypersurface $\ell_\mu x^\mu = 0$. 
	\item A lightray through this point, along which the foliation $\ell_\mu x^\mu = \const$ is arranged. 
	\item A 2d rotation angle around this lightray, which fixes the phase of the spinors $q^\alpha,\bar q^{\dot\alpha}$, and of the fields $\Phi_h$ with nonzero helicity.
\end{itemize}
Clearly, we can transform from any such frame to any other using de Sitter spacetime symmetries, i.e. the $t$-orthogonal components of \eqref{eq:P_action_real}-\eqref{eq:K_action_real}. However, as noted in section \ref{sec:extend:analytic_continuation}, there's a potential problem -- the inverse derivative in \eqref{eq:M_real} along the complex spatial vector $q^\alpha q^\beta\del_\beta{}^{\dot\alpha}t$. We will now show that one can transform from any lightcone frame to any other without activating this inverse derivative (and, in the chiral theory, without any inverse derivatives at all).

The challenge is to evolve from the lightcone of one bulk point (with a preferred lightray \& 2d rotation angle) to the lightcone of any other. This can be done by iterating two kinds of steps:
\begin{enumerate}
	\item Evolve from the given lightcone $\ell_\mu x^\mu = 0$ onto the lightcones of other points $\ell_\mu x^\mu = \const$ along the preferred lightray.
	\item Change the preferred lightray \& 2d rotation angle, through spacetime symmetries that \emph{preserve} the given lightcone $\ell_\mu x^\mu = 0$.
\end{enumerate}
In particular, we can transform between the lightcones of any two points $A,B$ that aren't lightlike-separated, by (a) evolving from $A$ along the preferred lightray to a point that's lightlike-separated from both $A$ and $B$, (b) re-orienting the preferred lightray to point towards $B$, and (c) evolving to $B$ along the new lightray. 

The first step -- evolution along the preferred lightray -- is just usual evolution using the lightcone-formalism field equations \eqref{eq:L_cubic_real}-\eqref{eq:field_eq_real}. In the original Metsaev formalism, the analogous operation would be to evolve with the Hamiltonian $P^-$. In the real theory of section \ref{sec:extend:real}, the field equations contain inverse derivatives along the real lightlike $\ell^\mu$, but not along the complex spatial $q^\alpha q^\beta\del_\beta{}^{\dot\alpha}t$. In the chiral theory + field frame \eqref{eq:L_cubic_chiral_again}-\eqref{eq:field_eq}, the field equations contain no inverse derivatives at all.

This leaves the second step -- spacetime symmetries that preserve the ``hyperplane'' (actually, lightcone) $\ell_\mu x^\mu = 0$, or equivalently, symmetries that preserve its origin point $(u,v,\mathbf{r})=(0,0,-\boldsymbol{\ell})$. In the original Metsaev formalism, the analogous operation would be to act with the kinematical generators $P^+,P^1,J^{+-},J^{+1},D,K^+,K^1$. In our case, the number of relevant generators is smaller, because the stabilizer group of a bulk origin point is lower than that of a boundary point. In fact, it is nothing but the Lorentz group at the bulk point! In Poincare coordinates, the conformal generators that preserve $\ell_\mu x^\mu = 0$ (and their subset that lies in the de Sitter group) are:
\begin{itemize}
	\item The $\ell^\mu$-orthogonal translations $q_\alpha P^{\alpha\dot\alpha},\bar q_{\dot\alpha} P^{\alpha\dot\alpha}$ (3 overall, 2 in the de Sitter group).
	\item The Lorentz components $q_\alpha J^{\alpha\beta}$ and $\bar q_{\dot\alpha} J^{\dot\alpha\dot\beta}$ (4 overall, 1 in the de Sitter group).
	\item Dilatations $D$ (1 overall, 1 in the de Sitter group).
	\item The $\ell^\mu$-orthogonal special conformals $q_\alpha K^{\alpha\dot\alpha},\bar q_{\dot\alpha} K^{\alpha\dot\alpha}$ (3 overall, 2 in the de Sitter group).
\end{itemize}
Let us evaluate these generators in the realization \eqref{eq:P_action_real}-\eqref{eq:M_real}, to see how they transform the fields on the lightcone $\ell_\mu x^\mu = 0$. It's easy to see that the interacting (i.e. second) piece of the intrinsic Lorentz \eqref{eq:M_real} never contributes, because it points along the $q^\alpha q^\beta$ and $\bar q^{\dot\alpha}\bar q^{\dot\beta}$ planes. Indeed, any would-be contribution involves a vanishing contraction of either $q^\alpha$ or $\bar q^{\dot\alpha}$ with itself. Thus, as in the original Metsaev formalism, the generators that preserve $\ell_\mu x^\mu = 0$ are \emph{kinematical}, in the sense that, when evaluated on $\ell_\mu x^\mu = 0$, they retain their free-field form. In particular, the inverse derivative along the complex $q^\alpha q^\beta\del_\beta{}^{\dot\alpha}t$ from \eqref{eq:M_real} never contributes. In fact, as in the original formalism, inverse derivatives don't appear in the kinematical generators on $\ell_\mu x^\mu = 0$ at all. Indeed, the inverse derivative in \eqref{eq:K_action_real} doesn't contribute to the kinematical components, while the one in the first term of \eqref{eq:M_real}, when it contributes, cancels against the derivative in the numerator.

To sum up, any two lightcone frames in our generalized formalism can be related by a combination of evolving with the equations of motion, and using symmetry generators that preserve the lightcone $\ell_\mu x^\mu = 0$. Neither of these steps involves inverse derivatives along complex directions. In the chiral theory + field frame, neither step involves inverse derivatives at all. In both the real and chiral theory, the lightcone-preserving generators are unaffected by interactions, and contain no inverse derivatives, i.e. their action on the fields on the fixed lightcone is local.

\section{Causality and covariance properties} \label{sec:causality}

In this section, we capitalize on our generalization of the lightcone formalism to present some causality properties. To our knowledge, this is the first treatment of causality in massless higher-spin interactions (for \emph{massive} higher-spin interactions, causality at the 4-point level famously imposes a string-theory-like structure \cite{Caron-Huot:2016icg,Kaplan:2020ldi,Kaplan:2020tdz}). Higher-spin theory aside, the lightcone formalism itself (despite its name!) is usually ill-suited for causality discussions, for two reasons:
\begin{itemize}
	\item The standard ``lightcone formalism'' is not about lightcones, but flat lightlike hyperplanes. There isn't enough flexibility in such foliations to probe interesting causal domains in spacetime.
	\item The formalism typically involves non-local inverse derivatives $1/(\ell\cdot\del)$ along the foliation's lightrays. 
\end{itemize}
As discussed in the previous section, our generalized lightcone formalism solves the first issue: we can now work with foliations involving general lightcones, not just flat hyperplanes. The second issue -- of non-locality along the lightrays -- remains. As a result, we should talk about causal domains of dependence for entire lightrays, not individual points. This turns out to be sufficient for some non-trivial causality properties, involving \emph{different lightcones that share a lightray}. In fact, we'll see that these properties are precisely what's needed to show that \emph{evolution in the de Sitter static patch} is causally consistent, i.e. is not contaminated by outside data. Our causal properties (all to leading order in the cubic interactions) can be summarized as follows:
\begin{enumerate}
	\item Consider solving the field equations in a given lightcone foliation, using perturbation theory with retarded propagators. When solving for the fields on one of the foliation's lightrays, the domain of dependence will fall inside the causal past of  \emph{any} lightcone that contains the given ray (including lightcones that aren't part of the foliation).
	\item Consider transforming between two lightcone foliations that share a lightray. On this shared ray, the transformation localizes: fields on the shared ray in the new foliation depend only on fields on the same ray in the old foliation. This has the side benefit of allowing us to define more covariant field quantities, which depend on the ray rather than on the entire lightcone frame.
\end{enumerate}
Let us expand on both these properties.

\subsection{Domains of dependence for the field equations} \label{sec:causality:domains}

In the lightcone formalism, one usually thinks of evolving the fields from one $\ell_\mu x^\mu = \const$ ``time slice'' of the foliation to the next. However, the field equations \eqref{eq:field_eq_real} can be studied in various spacetime regions, whose boundaries are not necessarily the $\ell_\mu x^\mu = \const$ slices of the foliation. That is the context of our 1st causal property. 

Consider, then, a given lightcone foliation, in which we solve the field equations \eqref{eq:field_eq_real} to the leading non-linear order in perturbation theory, using retarded propagators. In this setup, any violation of causality will be due to the interaction vertex, which can include arbitrarily many derivatives (as well as inverse derivatives along the foliation's lightrays, unless we're working with the chiral vertex in the chiral field frame). Now, suppose we're solving for the fields on one of the foliation's lightrays, e.g. the lightray $x^\mu\sim \ell^\mu$, which belongs to the slice $\ell_\mu x^\mu = 0$ of the foliation (note that the ray is truncated at the conformal boundary, in this case at $x^\mu=0$). By construction, the interaction vertices in the lightcone formalism only contain derivatives tangential to the $\ell_\mu x^\mu = \const$ slices. As a result, it's automatically true that the fields on the $x^\mu\sim \ell^\mu$ lightray are unaffected by data outside the causal past of the $\ell_\mu x^\mu = 0$ slice. But, as we learned in section \ref{sec:geometry}, the slice $\ell_\mu x^\mu = 0$ is nothing but the lightcone of a bulk point. It is then natural to wonder: what if we replace $\ell_\mu x^\mu = 0$ with any \emph{other} lightcone that includes the target ray $x^\mu\sim \ell^\mu$? The causal past of this lightcone may be smaller than that of $\ell_\mu x^\mu = 0$; for example, this is clearly true for the lightcone $x_\mu x^\mu = 0$ of the boundary point $x^\mu=0$. Thus, we ask: are the fields on the target ray $x^\mu\sim \ell^\mu$ unaffected by data outside the causal past of \emph{any} lightcone that contains it?

We argue that the answer is yes, due to the chiral nature of the cubic vertices. Indeed, the derivatives in the cubic vertex are not just tangential to the lightlike 3d ``time slice'' $\ell_\mu x^\mu = \const$, but more specifically to the totally-null 2d plane $q_\alpha x^{\alpha\dot\alpha} = \const$ (for chiral vertices), or $\bar q_{\dot\alpha} x^{\alpha\dot\alpha} = \const$ (for anti-chiral vertices). In particular, as we approach the target lightray $x^\mu\sim \ell^\mu$ from the past, the derivatives in the cubic vertex become tangential to one of the totally-null planes containing this ray: either the left-handed one $q_\alpha x^{\alpha\dot\alpha} = 0$, or the right-handed one $\bar q_{\dot\alpha} x^{\alpha\dot\alpha} = 0$. Now comes a handy geometric fact: in conformally-flat spacetimes like our (A)dS, any two lightcones that share a lightray will \emph{also} share the pair of totally-null planes that contain this ray. Therefore, as we approach the target lightray $x^\mu\sim \ell^\mu$ from the past, the derivatives inside the cubic vertex (no matter how many) become tangential to \emph{any} lightcone that contains it. This leads to the stated causal property of the perturbative evolution.

\subsection{Lightcone frames with a shared lightray} \label{sec:causality:shared}

We now turn to the 2nd causal property, which builds on observations made in \cite{Lang:2025rxt} for Higher-Spin self-dual GR. Consider transforming between two lightcone foliations that share a lightray, e.g. the ray $x^\mu\sim\ell^\mu$. Without loss of generality, we can focus on transformations that \emph{preserve} this ray (if instead it goes into a \emph{different} ray of the foliation, this can always be corrected by kinematical 2d translations). Generically in the lightcone formalism, the fields on the lightlike ``time slice'' containing $x^\mu\sim\ell^\mu$ in the new foliation must be expressed \emph{somehow} (specifically, by the symmetry generators) in terms of fields on the slice containing $x^\mu\sim\ell^\mu$ in the old foliation. Since all the lightrays on each slice are spacelike-separated from each other, \emph{causality} implies a restriction on this transformation: the fields on the fixed ray $x^\mu\sim\ell^\mu$ must transform among themselves, without any influence from data on other rays. As we'll now see, this causal property is indeed satisfied.

As with our analysis of lightcone-preserving transformations in section \ref{sec:geometry:transforming}, we can list the conformal generators that preserve the ray $x^\mu\sim\ell^\mu$, and their subsets that belong to the de Sitter group (or, alternatively, the AdS group with $\ell^\mu\del_\mu z\neq 0$):
\begin{itemize}
	\item Translations $\ell_\mu P^\mu$ along the lightray (1 overall, 0 in the de Sitter group).
	\item The Lorentz components $q_\alpha J^{\alpha\beta}$ and $\bar q_{\dot\alpha} J^{\dot\alpha\dot\beta}$ (4 overall, 1 in the de Sitter group).
	\item Dilatations $D$ (1 overall, 1 in the de Sitter group).
	\item Special conformals $K^\mu$ (4 overall, 3 in the de Sitter group).
\end{itemize}
As with the lightcone-preserving transformations from section \ref{sec:geometry:transforming}, when these generators are evaluated on the fixed ray $x^\mu\sim\ell^\mu$, two key simplifications occur. First, the interacting piece of the intrinsic Lorentz generators \eqref{eq:M_real}, which is along the planes $q^\alpha q^\beta$ and $\bar q^{\dot\alpha}\bar q^{\dot\beta}$, never contributes. Thus, the transformations of the fields on the fixed ray are purely kinematical. Second, whenever the free-field piece of the intrinsic Lorentz \eqref{eq:M_real} is non-vanishing, the derivatives in the numerator and denominator cancel. As a result, the transformations of the fields on the fixed ray consist only of the following:
\begin{enumerate}
	\item Moving from one point to another, according to the orbital part of the symmetry generators. These are local, i.e. without inverse derivatives. By construction, fields on the fixed ray always remain on it.
	\item Multiplication by real factors, due to dilatations of the local frame, described by the conformal-weight terms in $D$ and $K^\mu$. These act point-by-point, i.e. without derivatives at all.
	\item Multiplication by complex phases, due to 2d rotations of the local frame around the lightlike $\ell^\mu$, described by the helicity terms in $J^{\mu\nu}$ and $K^\mu$. These also act point-by-point.
	\item The intrinsic (last) term in the special conformals \eqref{eq:K_action_real}, which involves an integral over the lightray.
\end{enumerate}
None of these involve fields from outside the fixed lightray, so the causal property is upheld. Out of the above transformations on the fixed ray, the only non-local contribution is the intrinsic piece of the special conformals. As we've seen, in the chiral field frame, this piece is absent. Thus, in the chiral frame, the fields $\phi_h$ on the fixed ray undergo only orbital motions (along the ray) and complex rescalings (due to the local dilatations and Lorentz rotations). This holds not only in the chiral theory, but also in the presence of anti-chiral vertices: the interaction terms in the generators don't contribute either way.

\subsection{From lightcone fields to covariant quantities} \label{sec:causality:covariant}

The above discussion of transformations with a fixed ray has a useful upshot: in the chiral frame, the value of $\phi_h$ (as always, to leading order in the interactions) can be partially divorced from the lightcone foliation. Indeed, up to the complex rescalings from local dilatations and Lorentz, the value of $\phi_h$ is a property of only the point $x$ and the lightlike vector $\ell^\mu$. In fact, for negative helicities $h=-s$, the transformations of $\phi_h$ (which, we recall, are unaffected by interactions) are simply those of a linearized Weyl curvature component. 

To make this more concrete, let us drop the Poincare coordinates, and consider abstractly a point $x$, a lightlike vector $\ell^\mu$ at $x$, and its spinor square root $2\ell^\alpha\bar\ell^{\dot\alpha} = \ell^\mu e_\mu^{\alpha\dot\alpha}$. Here, $\ell^\alpha,\bar\ell^{\dot\alpha}$ are spinors in the internal flat tangent space, and $e_\mu^{\alpha\dot\alpha}$ is the (A)dS vielbein. Back in Poincare coordinates for e.g. de Sitter space, these are $e_\mu^{\alpha\dot\alpha} = \sigma_\mu^{\alpha\dot\alpha}/t$ and $\ell^\alpha = q^\alpha/\sqrt{t}$. Let's also fix the relative normalization of $\ell^\mu$ against the $t$ coordinate as:
\begin{align}
	\ell^\mu\del_\mu t = 1 \ . \label{eq:ell_t}
\end{align}
Then, at the free-field level, $\phi_{-s}$ is related to the gauge-invariant higher-spin Weyl curvature $\Psi_{\alpha_1\dots\alpha_{2s}}$ (with internal spinor indices) via:
\begin{align}
	-t\phi_{-s} = \ell^{\alpha_1}\dots\ell^{\alpha_{2s}}\Psi_{\alpha_1\dots\alpha_{2s}} \ . \label{eq:phi_Psi}
\end{align}
At the interacting level, it's no longer easy to define an invariant (or even covariant) Weyl curvature. However, we saw that the lightray-preserving transformations are unaffected by the interactions. Therefore, the behavior of $\phi_{-s}$ under local dilatations (which affect $t$ and the real scaling of $\ell^\alpha,\bar\ell^{\dot\alpha}$, via \eqref{eq:ell_t}) and local Lorentz (which affects the complex scaling of $\ell^\alpha,\bar\ell^{\dot\alpha}$) are the same as in the linearized case \eqref{eq:phi_Psi}. Furthermore, since the lightray-preserving transformations of $\phi_h$ are linear in the helicity $h$, we can deduce from \eqref{eq:phi_Psi} also the behavior of positive helicities $h>0$ (which, instead of Weyl curvature components, are related to lightcone-gauge prepotentials \cite{Lang:2025rxt}). Altogether, we get:
\begin{align}
	-t\phi_h \equiv \hat\phi_h(x^\mu;\ell^\alpha,\bar\ell^{\dot\alpha}) \ , \label{eq:hat_phi}
\end{align}
where $\hat\phi(x^\mu;\ell^\alpha)$ is a covariant quantity, in the sense that it depends only on the point $x^\mu$ and the internal-space spinors $\ell^\alpha,\bar\ell^{\dot\alpha}$ at that point. The behavior of $\hat\phi_h$ under complex rescaling of the spinors can be read off from \eqref{eq:phi_Psi} as:
\begin{align}
  \hat\phi_h(x^\mu;\rho\ell^\alpha,\bar\rho\bar\ell^{\dot\alpha}) = \rho^{-2h}\hat\phi_h(x^\mu;\ell^\alpha,\bar\ell^{\dot\alpha}) \ . \label{eq:hat_phi_scaling}
\end{align}
If we consider only chiral interactions, the whole argument can be repeated for lightcone frame transformations that preserve not a lightray (along $\ell^\mu$), but a totally-null left-handed plane (along $q^\alpha$, or equivalently $\ell^\alpha$). The list of generators from section \ref{sec:causality:shared} will now include \emph{all} the right-handed Lorentz components $J^{\dot\alpha\dot\beta}$, and $\hat\phi_h$ will end up depending only on $\ell^\alpha$, not on $\bar\ell^{\dot\alpha}$. Note that the result \eqref{eq:hat_phi}-\eqref{eq:hat_phi_scaling} does not involve approximations other than cutting off the (A)dS symmetry generators at cubic order. In particular, this means that \emph{if} the chiral theory is consistent to all orders with just cubic generators (as in the flat case), then \eqref{eq:hat_phi}-\eqref{eq:hat_phi_scaling} also holds to all orders in the chiral coupling.

Conversely, if we wish to treat both chiralities on an equal footing, we can equally well define \emph{anti-}chiral covariant quantities $\hat\phi^{\text{anti}}_h(x^\mu;\ell^\alpha,\bar\ell^{\dot\alpha})$ via:
\begin{align}
	-t\phi^{\text{anti}}_h \equiv \hat\phi^{\text{anti}}_h(x^\mu;\ell^\alpha,\bar\ell^{\dot\alpha}) \ ,
\end{align}
where $\phi^{\text{anti}}_h$ is defined by reversing eq. \eqref{eq:phi}:
\begin{align}
	\phi^{\text{anti}}_h = (\ell^\mu\del_\mu)^h\Phi_h = (\ell^\mu\del_\mu)^{2h}\phi_h \ . \label{eq:phi_anti}
\end{align}
To make the relationship between $\hat\phi_h$ and $\hat\phi^{\text{anti}}_h$ clearer, we can consider a lightcone foliation in which our point $x$ is near the ``lightlike infinity'' of the Poincare coordinates (which, we recall, is a perfectly regular hypersurface in de Sitter space -- a cosmological horizon). In this limit, $\ell^\mu$ becomes an \emph{affine} tangent vector along our lightray, while the $t$ factor in \eqref{eq:hat_phi},\eqref{eq:phi_anti} can be treated as a large constant. We then have:
\begin{align}
	\hat\phi^{\text{anti}}_h = (\ell^\mu\del_\mu)^{2h}\hat\phi_h \ , \label{eq:hat_phi_anti_relation}
\end{align}
which holds for affine $\ell^\mu$ (note that, having derived the relation \eqref{eq:hat_phi_anti_relation}, we can forget about the lightcone foliation that led to it). This allows us to define a \emph{non-chiral} covariant field quantity $\hat\Phi_h(x^\mu;\ell^\alpha,\bar\ell^{\dot\alpha})$ as:
\begin{align}
	\hat\Phi_h = (\ell^\mu\del_\mu)^h\hat\phi_h = (\ell^\mu\del_\mu)^{-h}\hat\phi^{\text{anti}}_h \ , \label{eq:hat_Phi}
\end{align}
where it's again important that we choose an affine $\ell^\mu$. Like the original lightcone field $\Phi_h$, the covariant quantity $\hat\Phi_h$ satisfies the reality condition $\hat\Phi_{-h} = \hat\Phi_h^\dagger$.

\section{Scattering in the static patch} \label{sec:static_patch}

We now turn to our main object of interest -- the scattering problem for HS gravity in the de Sitter static patch, to leading order in the interactions (i.e. 3-point scattering). This brings together the de Sitter lightcone formalism from section \ref{sec:extend}, its geometric interpretation from section \ref{sec:geometry}, and the causality and covariance properties from section \ref{sec:causality}. We will start with a maximally covariant statement of the scattering problem (section \ref{sec:static_patch:problem}), translate it into a lightcone-formalism computation in a Poincare patch (section \ref{sec:static_patch:lightcone}), then reformulate the latter in momentum and spinor-helicity variables (section \ref{sec:static_patch:momentum}). Finally, in section \ref{sec:static_patch:GR_like}, we'll present an example for interactions of the Self-Dual GR type, i.e. cubic vertices with $h_1+h_2+h_3=2$.

\subsection{Covariant problem statement} \label{sec:static_patch:problem}

Consider again the embedding-space picture of section \ref{sec:geometry:lightcones}: de Sitter space is the hyperboloid $x_Ix^I = -uv+\mathbf{r}^2 = 1$ in the 5d flat spacetime $\bbR^{1,4}$ with coordinates $x^I = (u,v,\mathbf{r})$ and metric $dx_I dx^I = -dudv + \mathbf{dr}^2$. We take both lightlike coordinates $u,v$ to be future-pointing. The boundaries of an observer's static patch can then be defined as:
\begin{itemize}
	\item Past horizon: $u=0,v<0,\mathbf{r}^2=1$.
	\item Future horizon: $u>0,v=0,\mathbf{r}^2=1$.
\end{itemize}
The horizons intersect at the 2-sphere $u=v=0$, known as the bifurcation sphere. We define a lightlike tangent $n^I=(0,2,\vec{0})$ for the past horizon, and $l^I = (2,0,\vec{0})$ for the future horizon. These vectors are covariantly constant along the respective horizons; in particular, they're affine along each of the horizons' lightrays. Along with the inner product $n_I l^I = -2$ on the bifurcation sphere, this property fixes $n^I$ and $l^I$ up to an overall constant rescaling $(n^I,\ell^I)\to(\rho^{-1}n^I,\rho\ell^I)$. We denote the spinor square roots of $n^I$ and $l^I$ (in the internal 4d tangent space, as in section \ref{sec:causality:covariant}) by $n^\alpha,\bar n^{\dot\alpha}$ and $l^\alpha,\bar l^{\dot\alpha}$. We define the phases of these spinors to be covariantly constant along their respective horizons' lightrays, and fix $n_\alpha l^\alpha = \bar n_{\dot\alpha} \bar l^{\dot\alpha} = 1$ on the bifurcation sphere. This leaves a phase freedom $(n^\alpha,l^\alpha,\bar n^{\dot\alpha},\bar l^{\dot\alpha})\to(e^{-i\theta}n^\alpha,e^{i\theta}l^\alpha,e^{i\theta}\bar n^{\dot\alpha},e^{-i\theta}\bar l^{\dot\alpha})$ at each point of the bifurcation sphere; these phases will be tied to the helicity phases of the spinning fields.

We can now invoke the covariant chiral field quantities $\hat\phi_h$ from section \ref{sec:causality:covariant}, and formulate the static-patch scattering problem as: \emph{express the final data $\hat\phi^{\text{out}}_h(x;l^\alpha,\bar l^{\dot\alpha})$ on the future horizon as a functional of the initial data $\hat\phi^{\text{in}}_h(x;n^\alpha,\bar n^{\dot\alpha})$ on the past horizon}. Here, the ``in/out'' labels are simply to clarify that the fields are evaluated at the past/future horizon. Alternatively, we can pose the problem in terms of the non-chiral data $\hat\Phi^{\text{out}}_h(x;l^\alpha,\bar l^{\dot\alpha})$ and $\hat\Phi^{\text{in}}_h(x;n^\alpha,\bar n^{\dot\alpha})$. Since $l^I$ and $n^I$ are affine, these are related to the chiral data as:
\begin{align}
	\hat\Phi^{\text{out}}_h(x;l^\alpha,\bar l^{\dot\alpha}) = (l^I\del_I)^h\hat\phi^{\text{out}}_h(x;l^\alpha,\bar l^{\dot\alpha}) \ ; \quad 
	\hat\Phi^{\text{in}}_h(x;n^\alpha,\bar n^{\dot\alpha}) = (n^I\del_I)^h\hat\phi^{\text{in}}_h(x;n^\alpha,\bar n^{\dot\alpha}) \label{eq:Phi_horizon}
\end{align}
In the following, we will work in terms of the chiral $\hat\phi_h$ for concreteness. 

We can conveniently combine the information of the spatial horizon position $\mathbf{r}$ and the spinors $n^\alpha,\bar n^{\dot\alpha}$ or $l^\alpha,\bar l^{\dot\alpha}$, by defining a spinor square root $\chi^\alpha,\bar\chi^{\dot\alpha}$ of $\mathbf{r}$ as:
\begin{align}
	\sigma_\mu^{\alpha\dot\alpha}\chi_\alpha\bar\chi_{\dot\alpha} = (1,\mathbf{r}) \ . \label{eq:chi}
\end{align}
Here, we conflate 3d and 4d spinors to avoid an excess of notation. We can now use the (arbitrary) complex phase of $\chi^\alpha$ to define the phases of $n^\alpha$ and $l^\alpha$. Specifically, we choose the phase of $l^\alpha$ to equal that of $\chi^\alpha$, which then fixes the phase of $n^\alpha$ such that $n_\alpha\chi^\alpha$ is real and positive (the proper geometric statement here is that the 4d totally-null bivector $l^\alpha l^\beta$ is the wedge product of $l^\mu$ with a real multiple of the 3d complex null vector $\chi^\alpha\chi^\beta$). With these conventions, the initial data $\hat\phi_h^{\text{in}}$ and final data $\hat\phi_h^{\text{out}}$ can be viewed as functions $\hat\phi_h^{\text{in}}(v;\chi^\alpha,\bar\chi^{\dot\alpha})$ and $\hat\phi_h^{\text{out}}(u;\chi^\alpha,\bar\chi^{\dot\alpha})$ of just the lightlike time $v$ or $u$ and the spinors $\chi^\alpha,\bar\chi^{\dot\alpha}$.  

Finally, we can subsume also the horizon lightlike time into spinor variables \cite{David:2019mos}. Specifically, we introduce spinors $\lambda^\alpha,\bar\lambda^{\dot\alpha}$ and $\mu^\alpha,\bar\mu^{\dot\alpha}$, whose direction and phase are the same as $\chi^\alpha,\bar\chi^{\dot\alpha}$, and whose magnitude is the Fourier transform of $v$ and $u$ respectively:
\begin{align}
	c_h^{\text{in}}(\lambda^\alpha,\pm\bar\lambda^{\dot\alpha}) &= \int_{-\infty}^\infty dv\,\hat\phi^{\text{in}}_h\!\left(v;\frac{\lambda^\alpha}{|\lambda|},\frac{\bar\lambda^{\dot\alpha}}{|\lambda|}\right)
	  e^{\pm i|\lambda|^2 v/2} \ ; \label{eq:c_in} \\
	c_h^{\text{out}}(\mu^\alpha,\pm\bar\mu^{\dot\alpha}) &= \int_{-\infty}^\infty du\,\hat\phi^{\text{out}}_h\!\left(u;\frac{\mu^\alpha}{|\mu|},\frac{\bar\mu^{\dot\alpha}}{|\mu|}\right)
      e^{\pm i|\mu|^2 u/2} \ . \label{eq:c_out}
\end{align}
Here, the $\pm$ signs describe positive/negative frequencies along the horizons' lightrays, and the spinors' magnitudes are defined as:
\begin{align}
  |\lambda| \equiv \sqrt{\sigma_t^{\alpha\dot\alpha}\lambda_\alpha\bar\lambda_{\dot\alpha}} \ ,
\end{align}
where we recall that $\sigma_t^{\alpha\dot\alpha}$ is the identity matrix. As we'll see more explicitly in section \ref{sec:static_patch:lightcone}, the spinors $\lambda^\alpha,\bar\lambda^{\dot\alpha}$ act as \emph{spinor-helicity variables} \cite{Maldacena:2011nz,David:2019mos} in the Poincare patch whose lightlike infinity is at the past horizon, and likewise for $\mu^\alpha,\bar\mu^{\dot\alpha}$ and the future horizon \cite{Albrychiewicz:2020ruh}. In terms of these spinor-helicity variables, the static-patch scattering problem can now be rephrased as: \emph{express the final data $c_h^{\text{out}}(\mu^\alpha,\pm\bar\mu^{\dot\alpha})$ as a functional of the initial data $c_h^{\text{in}}(\lambda^\alpha,\pm\bar\lambda^{\dot\alpha})$}. 

Note that in the Fourier transform \eqref{eq:c_in}-\eqref{eq:c_out}, we cover the entirety of the two horizons, rather than the two halves $v<0,u>0$ that bound the static patch. We're always able to consistently restrict to $v<0,u>0$ at the end of the calculation, thanks to \emph{causality}. This too will be discussed more explicitly in the next subsection.

\subsection{Computation scheme in the lightcone formalism} \label{sec:static_patch:lightcone}

With the static-patch scattering problem defined as above, we will now outline a computation procedure within the lightcone formalism, building on the lightcone-gauge constructions of \cite{Albrychiewicz:2021ndv,Neiman:2023bkq}. Along the way, we'll show that the causality properties from section \ref{sec:causality} are sufficient to make the static-patch problem well-defined (i.e. that the data on the $v<0$ past horizon is sufficient to determine the data on the $u>0$ future horizon). 

We start by choosing Poincare coordinates $x^\mu = (t,\mathbf{x})$ adapted to the past horizon, i.e. in which the past horizon appears as past lightlike infinity. These are just the coordinates defined by \eqref{eq:uv_from_Poincare}. The future horizon in these coordinates appears as $x_\mu x^\mu = 0$. We then choose a lightcone foliation defined by a lightlike vector $\ell^\mu = -\sigma^\mu_{\alpha\dot\alpha}q^\alpha\bar q^{\dot\alpha} = -e^\mu_{\alpha\dot\alpha}\ell^\alpha\bar\ell^{\dot\alpha}$, normalized via $\ell^\mu\del_\mu t = |q|^2 = 1$. We fix the phases of the spinors $q^\alpha,\ell^\alpha$ to match that of the future-horizon spinor $l^\alpha$ on the lightray $x^\mu\sim\ell^\mu$, which is shared between the future horizon and our lightcone foliation. We will now present a scheme for computing the final horizon data $\hat\phi^{\text{out}}_h(u;\chi^\alpha,\bar\chi^{\dot\alpha})$ \emph{on the shared ray $x^\mu\sim\ell^\mu$}, as a functional of the initial horizon data $\hat\phi^{\text{in}}_h(v;\chi^\alpha,\bar\chi^{\dot\alpha})$ on the past horizon $v<0$. By repeating this computation for every lightlike direction $\ell^\mu$, one can obtain the final data $\hat\phi^{\text{out}}_h$ on every lightray of the future horizon.

To see how this works, let's construct the mapping between the covariant horizon data from section \ref{sec:static_patch:problem} and the lightcone fields in our Poincare coordinates. We start with the data on the past horizon, which appears as past lightlike infinity in the Poincare coordinates, as discussed in section \ref{sec:geometry:lightcones}. Specifically, a point $x^I = (0,v,\mathbf{r})$ on the past horizon is described in Poincare coordinates by the limit:
\begin{align}
	t \to -\infty \ ; \quad \mathbf{x} = \left(-t + \frac{v}{2}\right)\mathbf{r} + \mathbf{x}_\perp \ , \label{eq:past_horizon_limit}
\end{align}
where $\mathbf{x}_\perp$ is orthogonal to $\mathbf{r}$ and finite. In this limit, the conversion from the Poincare-coordinate basis into the embedding-space basis sends any vector with finite components into a multiple of the past horizon's lightlike tangent $n^I = (0,2,\vec{0})$. In particular, this is true of $\ell^\mu$. Denoting $\ell^\mu = (1,\boldsymbol{\ell})$, we can find the proportionality coefficient between $\ell^\mu$ and $n^I$ as:
\begin{align}
	\frac{\ell^\mu\del_\mu v}{n^I\del_I v} = \frac{\ell^\mu\del_\mu(t - \mathbf{x}^2/t)}{2} = \frac{t^2 + \mathbf{x}^2 - 2t(\boldsymbol{\ell}\cdot\mathbf{x})}{2t^2} = 1 + \boldsymbol{\ell}\cdot\mathbf{r} 
	  = 2(q_\alpha\chi^\alpha)(\bar q_{\dot\alpha}\bar\chi^{\dot\alpha}) \ . \label{eq:ell_n}
\end{align}
Here, in the first equality we used the expression for $v$ from \eqref{eq:uv_from_Poincare}, in the third equality we used the limit \eqref{eq:past_horizon_limit}, and in the fourth equality we used the spinor square roots $-\sigma^\mu_{\alpha\dot\alpha}q^\alpha\bar q^{\dot\alpha} = (1,\boldsymbol{\ell})$ and \eqref{eq:chi}. Eq. \eqref{eq:ell_n} fixes for us the ratio of the vectors $\ell^\mu$ and $n^I$ on the past horizon. We can further fix the relative phase of their spinor square roots $\ell^\alpha,n^\alpha$ by comparing the phases of their inner products with $\chi^\alpha$. This fixes $\ell^\alpha$ w.r.t. $n^\alpha$ as:
\begin{align}
 \ell^\alpha = \sqrt{2}(q_\beta\chi^\beta)n^\alpha \ . 
\end{align}
Plugging this into \eqref{eq:hat_phi}-\eqref{eq:hat_phi_scaling}, we find the relation between the covariant initial data $\hat\phi^{\text{in}}_h(v;n^\alpha,\bar n^{\dot\alpha})$ and the asymptotic behavior of out lightcone field $\phi_h$ in the past-horizon limit \eqref{eq:past_horizon_limit}:
\begin{align}
	-t\phi_h = \frac{\hat\phi^{\text{in}}_h}{2^h(q_\alpha\chi^\alpha)^{2h}} \ . \label{eq:initial_data}
\end{align}
In particular, this implies that $-t\phi_h$ is finite in the past horizon limit \eqref{eq:past_horizon_limit}. This makes perfect sense, since the solutions $\phi_h$ to a massless field equation $\Box\phi_h = \dots$ typically decay as $\sim 1/t$ at lightlike infinity.

We can now use the lightcone field equations \eqref{eq:field_eq} or \eqref{eq:field_eq_real} to evolve the lightcone field $\phi_h$ from the initial data \eqref{eq:initial_data} into the future, and in particular onto the lightray $x^\mu\sim\ell^\mu$ of the future horizon. Here, the causal properties of section \ref{sec:causality} come into play. First, the causal property of section \ref{sec:causality:domains} ensures that the domain of dependence of the $x^\mu\sim\ell^\mu$ lightray is in the past not only of the lightcone $\ell_\mu x^\mu = 0$, but also of the future horizon $x_\mu x^\mu = 0$. In particular, this means that the initial data on the $v<0$ past horizon is sufficient to determine the fields $\phi_h$ on the $x^\mu\sim\ell^\mu$ lightray. We then invoke the causal property of section \ref{sec:causality:shared}, or equivalently the covariance property of section \ref{sec:causality:covariant}, to convert the values of $\phi_h$ on the $x^\mu\sim\ell^\mu$ ray in our lightcone foliation into the desired final data $\hat\phi^{\text{out}}_h$ defined w.r.t. the future horizon. 

In order to perform this conversion concretely, we will need the proportionality relation between $\ell^\alpha$ and $l^\alpha$ on the $x^\mu\sim\ell^\mu$ ray. From the embedding \eqref{eq:uv_from_Poincare}, we find the proportionality coefficient between our $\ell^\mu$ and the affine $l^I = (2,0,\vec{0})$ as: 
\begin{align}
	\frac{\ell^\mu\del_\mu u}{l^I\del_I u} = \frac{\ell^\mu\del_\mu(-1/t)}{2} = \frac{t^2}{2} \ . \label{eq:ell_l}
\end{align}
Recalling that we set the complex phases of $\ell^\alpha$ and $l^\alpha$ equal, we deduce the proportionality between $\ell^\alpha$ and $l^\alpha$ as:
\begin{align}
	\ell^\alpha = \frac{t}{\sqrt{2}}\,l^\alpha \ .
\end{align}
Plugging this into \eqref{eq:hat_phi}-\eqref{eq:hat_phi_scaling}, we obtain the conversion between the fields $\phi_h$ on the shared lightray and the covariant final data $\hat\phi^{\text{out}}_h(u;l^\alpha,\bar l^{\dot\alpha})$ as:
\begin{align}
	\hat\phi^{\text{out}}_h = -\frac{t^{2h+1}\phi_h}{2^h} \ . \label{eq:final_data}
\end{align}
This concludes our general procedure for solving the static-patch evolution problem using the field equations in the lightcone formalism. In the next subsection, we present the momentum-space version of this procedure.

\subsection{Computation scheme in momentum space with spinor-helicity variables} \label{sec:static_patch:momentum}

Let's now rewrite the above static-patch computation scheme in momentum space. We Fourier-transform the lightcone fields $\phi_h$ as:
\begin{align}
	\phi_h(x^\mu) = \int d^4k\,\tilde\phi_h(k_\mu)\,e^{ik_\mu x^\mu} \ , \label{eq:momentum_space}
\end{align} 
At the linearized level, the fields are solutions of the free field equation $\Box\phi_h=0$. These can be decomposed into plane waves whose 4-momentum $k_\mu$ is lightlike:
\begin{align}
	\tilde\phi_h^{(1)}(k_\mu) = \delta(k_\mu k^\mu)\,a_h(k_\mu) \ , \label{eq:initial_momenta}
\end{align}
At lightlike infinity of the Poincare coordinates, i.e. on the past horizon, these plane waves (when smeared slightly into wavepackets) exhibit a well-known asymptotic behavior:
\begin{itemize}
	\item They decay as $\sim 1/t$.
	\item They localize on a particular lightray of the past horizon, corresponding to the direction of $k^\mu$.
	\item Their frequency along this lightray becomes identified with $k^t = \pm |\mathbf{k}|$.
\end{itemize}
For details and a computation of the numerical factors involved, see e.g. \cite{Albrychiewicz:2020ruh}. Altogether, plugging in \eqref{eq:initial_data}, we find that the free-field momentum modes \eqref{eq:initial_momenta} are related to the spinor-helicity modes \eqref{eq:c_in} on the past horizon via: 
\begin{align}
	a_h(k_\mu) = \frac{|\lambda|^{2h}}{2^{h+2}\pi^2 i(q_\alpha\lambda^\alpha)^{2h}} \times \left\{\begin{array}{lc}
		c^{\text{in}}_h(\lambda^\alpha,\bar\lambda^{\dot\alpha}) & \quad k^t>0 \\  
		c^{\text{in}}_h(\lambda^\alpha,-\bar\lambda^{\dot\alpha}) & \quad k^t<0
	\end{array} \right. \ , \label{eq:initial_momenta_c}
\end{align}
where each (future-pointing or past-pointing) lightlike momentum $k^\mu$ is decomposed into spinors as: 
\begin{align}
	k^\mu = -\sign(k^t)\sigma^\mu_{\alpha\dot\alpha}\lambda^\alpha\bar\lambda^{\dot\alpha} \ . \label{eq:k}
\end{align}
At the interacting level, the momentum modes $\tilde\phi_h(k_\mu)$ receive higher-order corrections, and include non-lightlike momenta. Nevertheless, if we solve the lightcone field equations perturbatively using retarded propagators, the past horizon data $c^{\text{in}}_h$ will retain its linear relation \eqref{eq:initial_momenta}-\eqref{eq:initial_momenta_c} to the linearized field $\tilde\phi_h^{(1)}(k_\mu)$. 

Let us now turn to the final data $c^{\text{out}}_h$ on the lightray $x^\mu\sim\ell^\mu$ of the future horizon. Using eqs. \eqref{eq:uv_from_Poincare},\eqref{eq:c_out},\eqref{eq:final_data}-\eqref{eq:momentum_space}, this can be expressed as:
\begin{align}
  \begin{split}
	c_h^{\text{out}}(\mu^\alpha,\pm\bar\mu^{\dot\alpha}) &= \int_{-\infty}^\infty du\,\hat\phi^{\text{out}}_h\!\left(u;\frac{\mu^\alpha}{|\mu|},\frac{\bar\mu^{\dot\alpha}}{|\mu|}\right) e^{\pm i|\mu|^2 u/2} \\
	 &= \frac{1}{2^h}\int_{-\infty}^\infty \frac{du}{u^{2h+1}}\,\phi_h\!\left(x^\mu = \frac{\sigma^\mu_{\alpha\dot\alpha}\mu^\alpha\bar\mu^{\dot\alpha}}{|\mu|^2 u}\right) e^{\pm i|\mu|^2 u/2} \\
	 &= \frac{1}{2^h}\int d^4k\,\tilde\phi_h(k_\mu) \int_{-\infty}^\infty \frac{du}{u^{2h+1}}\,\exp\left(\frac{ik_{\alpha\dot\alpha}\mu^\alpha\bar\mu^{\dot\alpha}}{|\mu|^2 u} \pm \frac{i|\mu|^2 u}{2}\right) \\
	 &= \pm\frac{|\mu|^{4h}}{2^{3h}}\int d^4k\,\tilde\phi_h(k_\mu) \int_{-\infty}^\infty \frac{dU}{U^{2h+1}}\,\exp\left(iU \pm \frac{ik_{\alpha\dot\alpha}\mu^\alpha\bar\mu^{\dot\alpha}}{2U} \right) \ ,
  \end{split}
\end{align}
where we changed the integration variable as $U = \pm|\mu|^2u/2$. We can compute the $dU$ integral by closing it from above in the complex plane, and deforming the contour around the singularity at $U=0$. The deformation that leads to a well-defined answer is the one for which $\Im(\pm k_{\alpha\dot\alpha}\mu^\alpha\bar\mu^{\dot\alpha}/U)$ is positive. Thus, when $\pm k_{\alpha\dot\alpha}\mu^\alpha\bar\mu^{\dot\alpha}$ is positive (for lightlike or timelike $k^\mu$, this means that its energy sign is the same as that of $c_h^{\text{out}}$), we must bypass $U=0$ from below, and when it's negative, we must bypass from above. When we bypass from above, we get a closed contour with no singularities inside, so the integral vanishes. When we bypass from below, the contour becomes equivalent to a circle around $U=0$, and evaluates to a Bessel function of the first kind:
\begin{align}
  \oint_{-\infty}^\infty \frac{dU}{U^{2h+1}}\,\exp\left(iU \pm \frac{ik_{\alpha\dot\alpha}\mu^\alpha\bar\mu^{\dot\alpha}}{2U} \right) 
    = 2\pi i\left(\frac{-2}{\pm k_{\alpha\dot\alpha}\mu^\alpha\bar\mu^{\dot\alpha}}\right)^h\,J_{2h}\!\left(\sqrt{\pm 2k_{\alpha\dot\alpha}\mu^\alpha\bar\mu^{\dot\alpha}}\right) \ .
\end{align}
Thus, overall, the final data on the shared lightray is given by:
\begin{align}
	c_h^{\text{out}}(\mu^\alpha,\pm\bar\mu^{\dot\alpha}) = \pm\frac{2\pi i(-1)^h|\mu|^{4h}}{2^{2h}}\int d^4k\,\tilde\phi_h(k_\mu)\,\theta(\pm k_{\alpha\dot\alpha}\mu^\alpha\bar\mu^{\dot\alpha})\,
	  \frac{J_{2h}\!\left(\sqrt{\pm 2k_{\alpha\dot\alpha}\mu^\alpha\bar\mu^{\dot\alpha}}\right)}{(\pm k_{\alpha\dot\alpha}\mu^\alpha\bar\mu^{\dot\alpha})^h} \ . \label{eq:final_momenta}
\end{align}
Together, eqs. \eqref{eq:initial_momenta}-\eqref{eq:initial_momenta_c} and \eqref{eq:final_momenta} reduce the static patch scattering problem to the more standard problem of solving the lightcone field equations in momentum space.

\subsection{Example: gravity-like interactions} \label{sec:static_patch:GR_like}

For completeness, let us present an example of the equations of motion and their perturbative solution in momentum space (to leading order in the interactions). So far in this paper, the only vertices we wrote out explicitly in (A)dS were the Yang-Mills-like vertices with total helicity $h_1+h_2+h_3=1$, whose (A)dS expression coincides with the Minkowski one \eqref{eq:flat_vertex}, or equivalently \eqref{eq:V_YM_explicit}. Let us present here the GR-like vertices with $h_1+h_2+h_3=2$, for which the cosmological constant induces the simplest non-trivial modification of the Minkowski formula \eqref{eq:flat_vertex}. Instead of reading them off from Metsaev's expressions in \cite{Metsaev:2018xip}, we will use the lightcone-gauge formula \cite{Neiman:2024vit} for the cubic vertices of Higher-Spin Self-Dual GR \cite{Krasnov:2021nsq} (we performed the check by hand that the two agree up to integration by parts). HS Self-Dual GR is a chiral theory that features not quite all the cubic vertices with $h_1+h_2+h_3=2$, but those with one helicity negative and two positive. Conveniently, the negative-helicity field enters the Lagrangian as a multiplier (i.e. with no derivatives), so that the field equations for the positive-helicity fields can be read off immediately. Since the vertices with $h_1+h_2+h_3=2$ are linear in the helicities \cite{Metsaev:2018xip}, the field equations for the HS Self-Dual GR sector (one helicity negative, two positive) extend trivially into those for all vertices with $h_1+h_2+h_3=2$.

Focusing on the cubic vertex with a particular triple of helicities $(h_1,h_2,h_3)\equiv (h_1,h_2,-h)$ and omitting the coupling constant, the field equation from \cite{Neiman:2024vit} reads:
\begin{align}
  \begin{split}
	\Box\phi_h = q^\alpha q^\beta q^\gamma q^\delta&\left(-\frac{t}{2}\,\del_{\alpha\dot\alpha}\del_{\beta\dot\beta}\phi_{h_1}\del_\gamma{}^{\dot\alpha}\del_\delta{}^{\dot\beta}\phi_{h_2} 
        + (h_2-1)\,\del_{\alpha\dot\alpha}\del_{\beta\dot\beta}\phi_{h_1}\del_\gamma{}^{\dot\alpha}t\,\del_\delta{}^{\dot\beta}\phi_{h_2} \right. \\
        &\quad\qquad\qquad\qquad\qquad\qquad \left.\vphantom{\frac{t}{2}} {}+ (h_1-1)\,\del_{\alpha\dot\alpha}\del_{\beta\dot\beta}\phi_{h_2}\del_\gamma{}^{\dot\alpha}t\,\del_\delta{}^{\dot\beta}\phi_{h_1} \right) \ . 
   \end{split} \label{eq:GR_like_eq}
\end{align}
This equation can be solved using standard perturbation theory in momentum space. A nice feature is that, even though there's no translation symmetry along $t$, the perturbation theory still features delta functions w.r.t. the energy $k^t$. Indeed, $t$ appears in \eqref{eq:GR_like_eq} only as a linear factor, whose Fourier transform is a derivative acting on the $k^t$-preserving delta function. The situation is similar with all the chiral cubic vertices \eqref{eq:L_cubic_chiral} in the chiral field frame, whose dependence on $t$ is polynomial.

Let us now write the leading-order solution to \eqref{eq:GR_like_eq} explicitly. Starting from linearized momentum-space solutions \eqref{eq:initial_momenta} for $\phi_{h_1}$ and $\phi_{h_2}$, with spinor square roots for lightlike momenta as in \eqref{eq:k}, the quadratic solution for $\phi_h$ reads:
\begin{align}
  \begin{split}
	\tilde\phi^{(2)}_h(k_\mu) &= -\frac{8i(2\pi)^4}{k_\mu k^\mu}\int d^4k_1\,\delta(k_1\cdot k_1)\,a_{h_1}(k_1^\mu)\int d^4k_2\,\delta(k_2\cdot k_2)\,a_{h_2}(k_2^\mu)\,\delta^3(\mathbf{k}-\mathbf{k_1}-\mathbf{k_2}) \\
	  &\times \left(\langle q\lambda_1\rangle^2\langle q\lambda_2\rangle^2[\bar\lambda_1\bar\lambda_2]^2\frac{d}{dk^t} 
	   - (h_2-1)\langle q\lambda_1\rangle^2\langle q\lambda_2\rangle[\bar\lambda_1\bar\lambda_2](q^\alpha\bar\lambda_1^{\dot\alpha}\del_{\alpha\dot\alpha}t) \right. \\
	  &\qquad \left.\vphantom{\frac{d}{dk^2}} {}+ (h_1-1)\langle q\lambda_1\rangle\langle q\lambda_2\rangle^2[\bar\lambda_1\bar\lambda_2](q^\alpha\bar\lambda_2^{\dot\alpha}\del_{\alpha\dot\alpha}t) \right) \delta(k^t - k_1^t - k_2^t) \ ,
  \end{split} \label{eq:phi_solution}
\end{align}
where we used the notations $\langle q\lambda\rangle\equiv q_\alpha\lambda^\alpha$ and $[\bar q\bar\lambda]\equiv \bar q_{\dot\alpha}\lambda^{\dot\alpha}$ for inner products of spinors. The solution to the static-patch scattering problem is now given by plugging non-linear solutions of the form \eqref{eq:phi_solution} into the kinematical dictionary \eqref{eq:initial_momenta}-\eqref{eq:initial_momenta_c},\eqref{eq:final_momenta} between Poincare-patch and horizon modes. For more general choices of the interacting helicities, the detailed form of the field equation \eqref{eq:GR_like_eq} and the solution \eqref{eq:phi_solution} will become more complicated, featuring higher-order polynomials with respect to $t$, i.e. with respect to $d/dk^t$.

\section{Discussion} \label{sec:discuss}

In this paper, we set out to address the de Sitter static-patch scattering problem for cubic interactions in HS Gravity. For this purpose, we adapted the lightcone formalism for HS cubic vertices from AdS to de Sitter, by extending it to more general lightcone frames that employ bulk lightcones. This same generalization allowed us to formulate and verify some causality properties, novel for both HS theory and the lightcone formalism itself. As a side effect, one of the causality properties allowed us to convert the lightcone fields in a chiral field frame into covariant quantities, which only depend on a spacetime point and a lightlike vector (with spinor square roots) defined at that point. Together, the causality and covariance properties helped us formulate the static-patch scattering problem from within the lightcone formalism, show that it's causally consistent, and construct a perturbative computation scheme in coordinate space and in momentum space (using spinor-helicity variables). Along the way, we demonstrated that Yang-Mills-like cubic interactions of massless HS fields are conformally invariant, just like the cubic vertex of Yang-Mills theory itself.

Several future directions suggest themselves. First, it would be helpful to simplify (or just make more explicit) the polynomials $\calV,\calM$ that appear in the AdS lightcone cubic vertices of \cite{Metsaev:2018xip}. For this purpose, we point out the particular simplicity of the field equation \eqref{eq:GR_like_eq} in the GR-like sector $h_1+h_2+h_3=2$: the helicity-dependent terms depend on only one of the $h_i$'s at a time. This suggests that the vertices of \cite{Metsaev:2018xip} may be simplified when expressed in terms of the field equations, i.e. once we integrate by parts to strip all the derivatives from one of the fields.

Second, we should actually carry out the static-patch computation procedure of section \ref{sec:static_patch} for cubic interactions of arbitrary helicities. One will then seek to arrange the results for all the helicities in some pattern that exhibits HS symmetry. This would allow us to make contact between bulk static-patch processes and the holographic boundary CFT of \cite{Anninos:2011ui}, which would be a big step towards better understanding of holography in de Sitter space.

Third, there are still some gaps to fill in our construction of the static-patch calculation, involving boundary/edge issues. One should understand the role of edge modes on the bifurcation 2-sphere. Also, our treatment here was not specific enough to fix the integration constants for inverse derivatives that appear in the vertices \eqref{eq:L_cubic_real} in the non-chiral field frame, or, equivalently, in the conversion \eqref{eq:phi},\eqref{eq:hat_Phi}-\eqref{eq:Phi_horizon} between the chiral and non-chiral field frames. In our previous treatment of Self-Dual GR in \cite{Neiman:2023bkq}, the integration constants were fixed so as to keep the geometry at the bifurcation sphere undeformed, which ensured that the horizons maintain their constant area. A reasonable guess would be to generalize this rule to higher spins, but better understanding is needed. 

Finally, throughout this paper, we considered the geometry and causal structure of pure (A)dS, with the dynamical massless fields merely living on top of it. This is in line with the perspective \cite{Neiman:2015wma} on HS gravity as a theory of dynamical fields living on a fixed (A)dS geometry. However, in the original perspective \cite{Vasiliev:1990en,Vasiliev:1995dn,Vasiliev:1999ba}, HS gravity, like GR, is a diffeomorphism-invariant theory, with no pre-existing geometry aside from that defined by the dynamical fields. From this point of view, we should contend with deformations of the causal structure by the field perturbations: in particular, the spin-2 field should be deforming the spacetime metric. In the context of lightcone gauges and static-patch scattering, we addressed this challenge \cite{Neiman:2023bkq,Neiman:2024vit,Lang:2025rxt} for Self-Dual GR and its HS generalization. Doing the same for cubic interactions with arbitrary helicities may be an important next step.

\section*{Acknowledgements}

We are grateful to Mirian Tsulaia and Evgeny Skvortsov for discussions. This work was supported by the Quantum Gravity Unit of the Okinawa Institute of Science and Technology Graduate University (OIST). During part of the work, YN was visiting RIKEN iTHEMS.

\appendix
\section{Conformal symmetry of free fields and Yang-Mills-type interactions} \label{app:conformal}

In this Appendix, we demonstrate the full conformal symmetry of the free-field lightcone formalism for all helicities (section \ref{app:conformal:free}), and of the Yang-Mills-type cubic vertices with total helicity $h_1+h_2+h_3 = \pm 1$ (section \ref{app:conformal:cubic}). We will focus on the chiral case $h_1+h_2+h_3 = 1$. The anti-chiral case $h_1+h_2+h_3 = -1$ is analogous, and we work at leading order in the interactions, where the two don't mix. We work in the chiral field frame of section \ref{sec:extend:chiral}.

\subsection{Free fields} \label{app:conformal:free}

Here, we present our own derivation of the conformal symmetry of free massless fields in the lightcone formalism, proceeding as covariantly as possible, rather than component-by-component. The derivation is easiest in the language of section \ref{sec:extend:summary}, where the symmetry generators are defined in terms of their (linear) action on fields that satisfy the field equations. At the free-field level, the field equation is simply $\Box\phi_h = 0$, while the conformal generators can be copied from \eqref{eq:P_linear_chiral}-\eqref{eq:M_Delta_chiral} as:
\begin{align}
	\tilde P^{\alpha\dot\alpha}_{\text{lin.}} &= \del^{\alpha\dot\alpha} \ ; \label{eq:P_action_lin} \\
	\tilde J^{\alpha\beta}_{\text{lin.}} &= -x^{(\alpha}{}_{\dot\alpha}\del^{\beta)\dot\alpha} + \tilde M^{\alpha\beta}_{\text{lin.}} \ ; \label{eq:J_left_action_lin} \\
	\tilde J^{\dot\alpha\dot\beta}_{\text{lin.}} &= x^{\alpha(\dot\alpha}\del_\alpha{}^{\dot\beta)} \ ; \label{eq:J_right_action_lin} \\
	\tilde D_{\text{lin.}} &= -\frac{1}{2}x^{\alpha\dot\alpha} \del_{\alpha\dot\alpha} + \tilde\Delta \ ; \label{eq:D_action_lin}  \\ 
	\tilde K^{\alpha\dot\alpha}_{\text{lin.}} &= -\frac{1}{4}x_{\beta\dot\beta} x^{\beta\dot\beta} \del^{\alpha\dot\alpha} 
	- x^{\alpha\dot\alpha}\left(- \frac{1}{2}x^{\beta\dot\beta} \del_{\beta\dot\beta} + \tilde\Delta\right) - \frac{1}{2}x_\beta{}^{\dot\alpha}\tilde M^{\beta\alpha}_{\text{lin.}} \label{eq:K_action_lin} \ ,
\end{align}
where:
\begin{align}
	\tilde M^{\alpha\beta}_{\text{lin.}} = \frac{4h\bar q^{\dot\alpha} q^{(\alpha}\del^{\beta)}{}_{\dot\alpha}}{q^\gamma\bar q^{\dot\gamma}\del_{\gamma\dot\gamma}} \ ; \quad  \tilde\Delta = 1-h \ . 
\end{align}
Note that in this version, the Lorentz-violating lightcone nature of the formalism only shows up in the intrinsic Lorentz generator $\tilde M^{\alpha\beta}_{\text{lin.}}$. Without it, we're left with a manifestly Lorentz-covariant description of the free conformally-massless scalar.

To demonstrate the conformal symmetry, we must show that the generators \eqref{eq:P_action_lin}-\eqref{eq:K_action_lin} commute with the $\Box$ operator from the field equation, and that their commutators among themselves form the conformal algebra, up to terms of the form $(\dots)\Box$ that vanish on the field equations. And indeed, the commutators all take the required form. First, we have the easiest commutators, where $\tilde M^{\alpha\beta}_{\text{lin.}}$ either doesn't show up, or contributes trivially:
\begin{gather}
	[\tilde P^{\alpha\dot\alpha}_{\text{lin.}},\Box] = [\tilde J^{\alpha\beta}_{\text{lin.}},\Box] = [\tilde J^{\dot\alpha\dot\beta}_{\text{lin.}},\Box] = 0 \ ; \quad [\tilde D_{\text{lin.}},\Box] = -2\Box \ ; \\
	[\tilde P^{\alpha\dot\alpha}_{\text{lin.}},\tilde P^{\beta\dot\beta}_{\text{lin.}}] = 0 \ ; \quad [\tilde J^{\alpha\beta}_{\text{lin.}},\tilde P^{\gamma\dot\gamma}_{\text{lin.}}] = -2\epsilon^{\gamma(\alpha}\tilde P^{\beta)\dot\gamma}_{\text{lin.}} \ ; \quad 
	[\tilde J^{\dot\alpha\dot\beta}_{\text{lin.}},\tilde P^{\gamma\dot\gamma}_{\text{lin.}}] = 2\tilde P^{\gamma(\dot\alpha}_{\text{lin.}}\epsilon^{\dot\beta)\dot\gamma} \ ; \\
	[\tilde D_{\text{lin.}},\tilde P^{\alpha\dot\alpha}_{\text{lin.}}] = -\tilde P^{\alpha\dot\alpha}_{\text{lin.}} \ ; \quad [\tilde D_{\text{lin.}},\tilde J^{\alpha\beta}_{\text{lin.}}] = [\tilde D_{\text{lin.}},\tilde J^{\dot\alpha\dot\beta}_{\text{lin.}}] = 0 \ ; \quad 
	[\tilde D_{\text{lin.}},\tilde K^{\alpha\dot\alpha}_{\text{lin.}}] = \tilde K^{\alpha\dot\alpha}_{\text{lin.}} \ ; \label{eq:D_commutators} \\
	[\tilde J^{\dot\alpha\dot\beta}_{\text{lin.}},\tilde J^{\dot\gamma\dot\delta}_{\text{lin.}}] = 2\left(\epsilon^{\dot\alpha(\dot\gamma}\tilde J_{\text{lin.}}^{\dot\delta)\dot\beta} + \epsilon^{\dot\beta(\dot\gamma}\tilde J_{\text{lin.}}^{\dot\delta)\dot\alpha}\right) \label{eq:J_R_J_R} \ .
\end{gather}
Next, there are commutators where the contribution of $\tilde M^{\alpha\beta}_{\text{lin.}}$ requires some calculation, including the use of Fierz identities:
\begin{align}
	[\tilde K^{\alpha\dot\alpha}_{\text{lin.}},\Box] &= 2\left(x^{\alpha\dot\alpha} + \frac{2hq^\alpha\bar q^{\dot\alpha}}{q^\beta\bar q^{\dot\beta}\del_{\beta\dot\beta}} \right)\Box \ ; \\
	[\tilde P^{\alpha\dot\alpha}_{\text{lin.}},\tilde K^{\beta\dot\beta}_{\text{lin.}}] &= \epsilon^{\dot\alpha\dot\beta}\tilde J^{\alpha\beta}_{\text{lin.}} + \epsilon^{\alpha\beta}\tilde J^{\dot\alpha\dot\beta}_{\text{lin.}} 
       + 2\epsilon^{\alpha\beta}\epsilon^{\dot\alpha\dot\beta}\tilde D_{\text{lin.}} \ ; \label{eq:P_K} \\
	[\tilde J^{\alpha\beta}_{\text{lin.}},\tilde J^{\dot\alpha\dot\beta}_{\text{lin.}}] &= \frac{8hq^\alpha q^\beta\bar q^{\dot\alpha}\bar q^{\dot\beta}}{(q^\gamma\bar q^{\dot\gamma}\del_{\gamma\dot\gamma})^2}\Box \ ; \label{eq:J_L_J_R} \\
	[\tilde J^{\dot\alpha\dot\beta}_{\text{lin.}},\tilde K^{\gamma\dot\gamma}_{\text{lin.}}] &= 2\tilde K^{\gamma(\dot\alpha}_{\text{lin.}}\epsilon^{\dot\beta)\dot\gamma} 
	   + \frac{4h\bar q^{\dot\alpha}\bar q^{\dot\beta}q^\gamma q^\alpha x_\alpha{}^{\dot\gamma}}{(q^\delta\bar q^{\dot\delta}\del_{\delta\dot\delta})^2}\Box \ , \label{eq:J_R_K}
\end{align}
where to compute \eqref{eq:J_L_J_R}-\eqref{eq:J_R_K}, it helps to first establish:
\begin{align}
	[\tilde M^{\alpha\beta}_{\text{lin.}},x^{\gamma\dot\gamma}] = \frac{8hq^\alpha q^\beta \bar q^{\dot\gamma}\bar q^{\dot\alpha}\del^\gamma{}_{\dot\alpha}}{(q^\delta\bar q^{\dot\delta}\del_{\delta\dot\delta})^2} \ .
\end{align}
The remaining commutators $[\tilde J^{\alpha\beta}_{\text{lin.}},\tilde J^{\gamma\delta}_{\text{lin.}}]$, $[\tilde J^{\alpha\beta}_{\text{lin.}},\tilde K^{\gamma\dot\gamma}_{\text{lin.}}]$ and $[\tilde K^{\alpha\dot\alpha}_{\text{lin.}},\tilde K^{\beta\dot\beta}_{\text{lin.}}]$ are harder. For $[\tilde J^{\alpha\beta}_{\text{lin.}},\tilde J^{\gamma\delta}_{\text{lin.}}]$ and $[\tilde J^{\alpha\beta}_{\text{lin.}},\tilde K^{\gamma\dot\gamma}_{\text{lin.}}]$, we can use a trick: instead of our chiral field frame \eqref{eq:phi}, we switch to the \emph{anti-chiral} field frame \eqref{eq:phi_anti}. We can then read off $[\tilde J^{\alpha\beta}_{\text{lin.}},\tilde J^{\gamma\delta}_{\text{lin.}}]$ and $[\tilde J^{\alpha\beta}_{\text{lin.}},\tilde K^{\gamma\dot\gamma}_{\text{lin.}}]$ as the mirror images of \eqref{eq:J_R_J_R} and \eqref{eq:J_R_K}, with left-handed and right-handed indices interchanged and helicities flipped as $h\to -h$ (up to conjugating by the transformation \eqref{eq:phi_anti}, which turns out to be trivial):
\begin{align}
	[\tilde J^{\alpha\beta}_{\text{lin.}},\tilde J^{\gamma\delta}_{\text{lin.}}] &= 2\left(\epsilon^{\alpha(\gamma}\tilde J_{\text{lin.}}^{\delta)\beta} + \epsilon^{\beta(\gamma}\tilde J_{\text{lin.}}^{\delta)\alpha}\right) \ ; \\
	[\tilde J^{\alpha\beta}_{\text{lin.}},\tilde K^{\gamma\dot\gamma}_{\text{lin.}}] &= -2\epsilon^{\gamma(\alpha}\tilde K^{\beta)\dot\gamma}_{\text{lin.}}
	  - \frac{4h q^\alpha q^\beta\bar q^{\dot\gamma}\bar q^{\dot\alpha} x^\gamma{}_{\dot\alpha}}{(q^\delta\bar q^{\dot\delta}\del_{\delta\dot\delta})^2}\Box \ . \label{eq:J_L_K}
\end{align}
This leaves the most difficult commutator $[\tilde K^{\alpha\dot\alpha}_{\text{lin.}},\tilde K^{\beta\dot\beta}_{\text{lin.}}]$, which we can handle indirectly. First, we can use the Jacobi identity and the known commutators of $\tilde K^{\alpha\dot\alpha}_{\text{lin.}}$ with the other generators to conclude that $\left[\del^{\gamma\dot\gamma}, [\tilde K^{\alpha\dot\alpha}_{\text{lin.}},\tilde K^{\beta\dot\beta}_{\text{lin.}}]\right]$ vanishes on the equations of motion:
\begin{align}
	\left[\del^{\gamma\dot\gamma}, [\tilde K^{\alpha\dot\alpha}_{\text{lin.}},\tilde K^{\beta\dot\beta}_{\text{lin.}}]\right] 
	= \left[\tilde P^{\gamma\dot\gamma}_{\text{lin.}}, [\tilde K^{\alpha\dot\alpha}_{\text{lin.}},\tilde K^{\beta\dot\beta}_{\text{lin.}}]\right] = (\dots)\Box \ .
\end{align}
This means that $[\tilde K^{\alpha\dot\alpha}_{\text{lin.}},\tilde K^{\beta\dot\beta}_{\text{lin.}}]$ itself also vanishes on the equations of motion, \emph{up to an $x^\mu$-independent piece}. But such an $x^\mu$-independent piece cannot arise. Indeed, all terms in $\tilde K^{\alpha\dot\alpha}_{\text{lin.}}$ are of the form $f(x)g(\del)$, where $f(x)$ are polynomials with strictly positive powers of $x^\mu$. The commutator of any two such terms is itself of the same form, and thus can never produce an $x^\mu$-independent piece. We conclude that $[\tilde K^{\alpha\dot\alpha}_{\text{lin.}},\tilde K^{\beta\dot\beta}_{\text{lin.}}]$ vanishes on the equations of motion:
\begin{gather}
	[\tilde K^{\alpha\dot\alpha}_{\text{lin.}},\tilde K^{\beta\dot\beta}_{\text{lin.}}] = (\dots)\Box \ .
\end{gather}
This concludes our derivation of conformal symmetry for the free-field lightcone formalism.

\subsection{Yang-Mills-like cubic interactions} \label{app:conformal:cubic}

We now turn to establish the conformal invariance (to leading order in the interactions) of the cubic vertices \eqref{eq:flat_vertex}-\eqref{eq:flat_M} with $h_1+h_2+h_3=1$. Note that for these vertices, the AdS formulas \eqref{eq:AdS_vertex}-\eqref{eq:Q_M_limit} coincide with the flat ones, and in particular don't contain any factors of the $z$ coordinate. 

Here, we will work in Metsaev's component formalism. The coordinates are $x^\mu = (x^+,x^-,x^1,x^2)$, with metric $ds^2 = 2dx^+dx^1 + (dx^1)^2 + (dx^2)^2$. In these coordinates, we write the preferred lightlike vector as $\ell^\mu = (0,1,0,0)$, i.e.:
\begin{align}
	\ell_\mu x^\mu = x^+ \ ; \quad \ell_\mu\del^\mu = \del^+ \ . \label{eq:ell_components}
\end{align}
Derivatives $p^{(i)}_{\dot\alpha}$ of the $i$'th field along the left-handed plane $q^\alpha q^\beta$, which appear in the interactions, can be packaged as:
\begin{align}
	p^{(i)}_{\dot\alpha}p^{(j)\dot\alpha} = q^\alpha q^\beta\del^{(i)}_{\alpha\dot\alpha} \del^{(j)}_\beta{}^{\dot\alpha} = \del^+_{(i)}\big(\del_{(j)}^2 + i\del_{(j)}^1\big) - \big((i)\leftrightarrow (j)\big) \ . \label{eq:q_components}
\end{align}
The generators $P^+,P^1,P^2,J^{+-},J^{+1},J^{+2},J^{12},D,K^+,K^1,K^2$, which preserve the hyperplane $x^+=0$, are ``kinematical'', i.e. unaffected by the interactions. The interactions enter into the other, ``dynamical'', generators $P^-,J^{-1},J^{-2},K^-$. From existing results, we know that the following subalgebras form symmetries of the action with the correct commutators:
\begin{itemize}
	\item The algebra of kinematical generators $(P^+,P^1,P^2,J^{+-},J^{+1},J^{+2},J^{12},D,K^+,K^1,K^2)$, since these are the same as in the free theory, whose conformal symmetry we just established.
	\item The Poincare algebra $(P^+,P^-,P^1,P^2,J^{+-},J^{+1},J^{+2},J^{-1},J^{-2},J^{12})$, since we know that the $h_1+h_2+h_3=1$ vertices are Poincare-invariant \cite{Bengtsson:1983pd,Fradkin:1991iy}.
	\item The AdS algebra $(P^+,P^-,P^1,J^{+-},J^{+1},J^{-1},D,K^+,K^-,K^1)$, since we know that the same $h_1+h_2+h_3=1$ vertices are also AdS-invariant \cite{Metsaev:2018xip}.
	\item The ``other'' AdS algebra $(P^+,P^-,P^2,J^{+-},J^{+2},J^{-2},D,K^+,K^-,K^2)$, which is equally good, since the $h_1+h_2+h_3=1$ vertices don't single out either of the $x^1,x^2$ axes as the ``special'' $z$ axis.
\end{itemize}
Together, these subalgebras establish all the would-be conformal generators as symmetries of the action, and guarantee the correctness of almost all commutators, except $[K^2,J^{-1}]$, $[K^1,J^{-2}]$, and $[J^{12},K^-]$. Of these, the first two are completely analogous, and the third can be reduced to them via $J^{12} = [P^1,K^2]$ and the Jacobi identity. Thus, to establish conformal symmetry, we only need to check one commutator, $[K^2,J^{-1}]$, which should vanish. More specifically, since at the free-field level the conformal algebra is guaranteed, and $K^2$ is kinematical, we only need to check that the \emph{free-field} $K^2$ commutes with the \emph{interaction term} in $J^{-1}$:
\begin{align}
	\big[K^2_{[2]},J^{-1}_{[3]}\big] = 0 \ .
\end{align}
Instead of working with the quadratic QFT operator $K^2_{[2]}$, it's more convenient to directly use its linear action on the fields. Working in the chiral field frame, this reads:
\begin{align}
	\tilde K^2_{\text{lin.}} = \frac{1}{2}\left((x^1)^2 + (x^2)^2\right)\del^2 - x^2(x^-\del^+ + x^1\del^1 + x^2\del^2 + 1) + h(x^2 - ix^1) \ . \label{eq:K_2}
\end{align}
Let us now write $J^{-1}_{[3]}$. Following Metsaev and our \eqref{eq:P_general}-\eqref{eq:J_general}, we will express it in terms of fields and their derivatives $\del^+,\del^1,\del^2$ along the initial hyperplane $x^+ = 0$. Plugging in \eqref{eq:M_R_cubic},\eqref{eq:L_cubic_chiral}-\eqref{eq:M_cubic_chiral}, focusing on a single set of helicities $h_1,h_2,h_3$ and dropping the coupling constant, we get:
\begin{align}
	J^{-1}_{[3]} =  \int d^3x\left(-x^1\calV_{h_1h_2h_3} + i\calM_{h_1h_2h_3}\right)\phi_{h_1}\phi_{h_2}\phi_{h_3} \ . \label{eq:J_-1}
\end{align}
Here, the vertices $\calV_{h_1h_2h_3},\calM_{h_1h_2h_3}$ for $h_1+h_2+h_3=1$ can be read off from \eqref{eq:P_beta},\eqref{eq:flat_vertex}-\eqref{eq:flat_M},\eqref{eq:ell_components}-\eqref{eq:q_components} as:
\begin{align}
	\calV_{h_1h_2h_3} &= \del^+_{(1)}\big(\del_{(2)}^2 + i\del_{(2)}^1\big) - \del^+_{(2)}\big(\del_{(1)}^2 + i\del_{(1)}^1\big) + \text{cyclic permutations} \ ; \label{eq:V_YM_explicit} \\
	\calM_{h_1h_2h_3} &= 2(h_2-h_3)\del^+_{(1)} + \text{cyclic permutations} \ ,
\end{align}
where we dropped an overall factor of $1/3$, and the cyclic permutations are over the field labels $i=1,2,3$. With these ingredients, we can write the desired commutator as the action of the linearized generator \eqref{eq:K_2} on the fields inside \eqref{eq:J_-1}:
\begin{align}
	\big[K^2_{[2]},J^{-1}_{[3]}\big] = \int d^3x\left(-x^1\calV_{h_1h_2h_3} + i\calM_{h_1h_2h_3}\right)\big(\tilde K^2_{\text{lin.}(1)} + \tilde K^2_{\text{lin.}(2)} + \tilde K^2_{\text{lin.}(3)}\big)\phi_{h_1}\phi_{h_2}\phi_{h_3} \ . \label{eq:commutator}
\end{align}
Here, in each of the operators $\tilde K^2_{\text{lin.}(i)}$, the helicity, position and derivatives $h,x^\mu,\del^\mu$ in \eqref{eq:K_2} are given labels $h_i,x^\mu_{(i)},\del^\mu_{(i)}$ referring to the $i$'th field $\phi_{h_i}$. After all the derivatives $\del^\mu_{(i)}$ inside \eqref{eq:commutator} act, we set all the $x^\mu_{(i)}$'s to the same value $x^\mu$, which is then integrated over the $x^+=0$ hyperplane with the measure $d^3x = dx^-dx^1dx^2$. 

With these preliminaries, the commutator \eqref{eq:commutator} can be readily evaluated. The main idea is to commute all the labeled coordinates $x^\mu_{(i)}$ to the left of the labeled derivatives $\del^\mu_{(i)}$. Once this is done, we can drop the coordinates' labels $x^\mu_{(i)}\to x^\mu$, and we can combine labeled derivatives as $\del^\mu_{(1)}+\del^\mu_{(2)}+\del^\mu_{(3)}\to \del^\mu$. In more detail, the calculation can be done in the following steps:
\begin{enumerate}
	\item Commute the $\tilde K^2_{\text{lin.}(1)} + \tilde K^2_{\text{lin.}(2)} + \tilde K^2_{\text{lin.}(3)}$ to the left of the $\calV_{h_1h_2h_3}$ and $\calM_{h_1h_2h_3}$. In the extra terms that arise from this commutation, commute all the coordinates to the left of the derivatives. 
	\item Once the $\tilde K^2_{\text{lin.}(1)} + \tilde K^2_{\text{lin.}(2)} + \tilde K^2_{\text{lin.}(3)}$ is on the left of $\calV_{h_1h_2h_3}$ and $\calM_{h_1h_2h_3}$, the terms with derivatives inside $\tilde K^2_{\text{lin.}(1)} + \tilde K^2_{\text{lin.}(2)} + \tilde K^2_{\text{lin.}(3)}$ can all be combined using $x^\mu_{(i)}\to x^\mu$ and $\del^\mu_{(1)}+\del^\mu_{(2)}+\del^\mu_{(3)}\to \del^\mu$. Commute the resulting $\del^\mu$ to the left, where it can become a total derivative and vanish upon integration.
	\item Combine helicity factors using $h_1+h_2+h_3=1$.
\end{enumerate}
After these steps, the $-x^1\calV_{h_1h_2h_3}$ and $i\calM_{h_1h_2h_3}$ terms in \eqref{eq:commutator} evaluate to $ix^1\calM_{h_1h_2h_3}$ and $-ix^1\calM_{h_1h_2h_3}$, which cancel, giving the desired result $[K^2,J^{-1}] = 0$.

\section{Comparing coupling constants between the Fronsdal and lightcone formalisms} \label{app:matching}

In this Appendix, we derive the scaling constant $a$ in the proportionality \eqref{eq:c_proportionality} between the chiral cubic couplings of different helicities. We do this for the case of type-A HS gravity, by comparing to its holographic dual -- the free $O(N)$ vector model. It is sufficient to consider vertices of the form $(h_1,h_2,h_3)=(0,0,\pm s)$, and to focus on the highest-derivative terms (equivalently, the flat limit, or the large-$z$ limit). 

Let us start by writing the most general real lightcone Lagrangian to cubic order, with its chiral couplings subject to the pattern \eqref{eq:c_proportionality} derived from the chiral sector's quartic consistency. Using \eqref{eq:action},\eqref{eq:p}-\eqref{eq:L_cubic},\eqref{eq:c_proportionality}-\eqref{eq:Q_V_limit}, this can be written as:
\begin{align}
	\calL ={}& \frac{1}{2}\Phi_0\Box\Phi_0 + \sum_{s>0}\Phi_{-s}\Box\Phi_s \label{eq:lightcone_00s_raw} \\
	&+ \sum_{s>0}\frac{z^{s-1}}{(s-1)!}\,\Phi_0\left(Ca^s  \big(q^{\alpha_1}\del_{\alpha_1\dot\alpha_1}\dots q^{\alpha_s}\del_{\alpha_s\dot\alpha_s}\Phi_0\big)
	\left(\frac{q^{\beta_1}\del_{\beta_1}{}^{\dot\alpha_1}\dots q^{\beta_s}\del_{\beta_s}{}^{\dot\alpha_s}}{(\ell\cdot\del)^s}\,\Phi_s\right) + c.c. \right) \ , \nonumber
\end{align}
where $C$ and $a$ are (for the moment) complex constants, and we display only the vertices with $(h_1,h_2,h_3)=(0,0,\pm s)$. The fields $\Phi_h$ ($h=0,\pm s$) satisfy the reality conditions $\Phi_h^\dagger = \Phi_{-h}$. When writing the vertices in \eqref{eq:lightcone_00s_raw}, we used the flat limit (i.e. neglected terms where the derivatives act on the $z$ factor) to replace $\bbP\to p^{(2)}_{\dot\alpha}p^{(3)\dot\alpha}$ via integration-by-parts, making the first $\Phi_0$ factor derivative-free. 

Now, in the type-A theory, the overall chiral coupling $C$ is real \cite{Skvortsov:2018uru}, and can always be made positive by flipping the sign of all the fields $\Phi_h\to -\Phi_h$. Furthermore, as discussed in the main text, we can use phase rescalings $\Phi_h\to e^{ih\theta}\Phi_h$ to make $a$ real and positive as well. We can then take $C$ and $a$ outside the ``${}+c.c.$'' parentheses. Writing out both chiral and anti-chiral vertices explicitly, we obtain:
\begin{align}
   \calL ={}& \frac{1}{2}\Phi_0\Box\Phi_0 + \sum_{s>0}\Phi_{-s}\Box\Phi_s \label{eq:lightcone_00s} \\
   	    &+ C\sum_{s>0}\frac{a^sz^{s-1}}{(s-1)!}\,\Phi_0\left(  \big(q^{\alpha_1}\del_{\alpha_1\dot\alpha_1}\dots q^{\alpha_s}\del_{\alpha_s\dot\alpha_s}\Phi_0\big)
   	    \left(\frac{q^{\beta_1}\del_{\beta_1}{}^{\dot\alpha_1}\dots q^{\beta_s}\del_{\beta_s}{}^{\dot\alpha_s}}{(\ell\cdot\del)^s}\,\Phi_s\right) \right. \nonumber \\
   	    &\qquad\qquad\qquad\qquad\left.{} + \big(\bar q^{\dot\alpha_1}\del_{\alpha_1\dot\alpha_1}\dots\bar q^{\dot\alpha_s}\del_{\alpha_s\dot\alpha_s}\Phi_0\big) \
   	    \left(\frac{\bar q^{\dot\beta_1}\del^{\alpha_1}{}_{\dot\beta_1}\dots\bar q^{\dot\beta_s}\del^{\alpha_s}{}_{\dot\beta_s}}{(\ell\cdot\del)^s}\,\Phi_{-s} \right) \right) \ . \nonumber
\end{align}
Before comparing to the holographic dual, let us emphasize the conceptual status of the Lagrangian \eqref{eq:lightcone_00s}. First, note that the real Lagrangian \eqref{eq:lightcone_00s} can be considered as-is, irrespective of its origin from the chiral theory. This makes it clear that different positive values of $a$ are indeed distinct: there is no further rescaling of the fields that can change one Lagrangian \eqref{eq:lightcone_00s} into another with a different positive $a$. For each of these Lagrangians, the chiral sector's quartic consistency conditions \emph{in the flat limit} are satisfied. On the other hand, in the holographic CFT dual, there is no 1-parameter freedom that would correspond to different choices of $a$. This implies that quartic consistency does, after all, require a particular value of $a$, but that this stricter consistency condition cannot be accessed from just the flat limit: the full AdS consistency analysis would probably need to be carried out.

With this understanding, we are ready to read off the correct value of $a$ from the simplest holographic dual -- the free $O(N)$ vector model. To perform the comparison, we will need the covariant version of the Lagrangian \eqref{eq:lightcone_00s}. This was worked out in \cite{Bekaert:2015tva} (and for all spins $s_1,s_2,s_3$ in \cite{Sleight:2016dba}), using the holographic correlators to fix the cubic couplings. The field variables in \cite{Bekaert:2015tva} are a scalar $\varphi$ and spin-$s$ Fronsdal fields $\varphi_{\mu_1\dots\mu_s}$ \cite{Fronsdal:1978vb}, which we take to be transverse and traceless. Since in this paper we raise/lower indices with the flat metric $\eta_{\mu\nu}$, we will be careful to define $\varphi_{\mu_1\dots\mu_s}$ with lower indices, and to include explicit powers of $g^{\mu\nu} = z^2\eta^{\mu\nu}$ when its indices are raised. We also include in the Lagrangian an explicit factor of $\sqrt{-g}=1/z^4$. On the other hand, we use the flat limit to write AdS covariant derivatives simply as partial derivatives $\del_\mu$. With these settings, the cubic Lagrangian of \cite{Bekaert:2015tva,Sleight:2016dba} reads:
\begin{align}
	\begin{split}
		\calL ={}& \frac{1}{2z^2}\Big(\varphi\Box\varphi + \sum_{s>0} z^{2s}\eta^{\mu_1\nu_1}\dots\eta^{\mu_s\nu_s}\varphi_{\mu_1\dots\mu_s}\Box\varphi_{\nu_1\dots\nu_s}\Big) \\
		&+ C'\sum_{s>0} \frac{2^{s/2}z^{2s-4}}{(s-1)!}\,\eta^{\mu_1\nu1}\dots\eta^{\mu_s\nu_s}\varphi\,\varphi_{\mu_1\dots\mu_s}\del_{\nu_1}\dots\del_{\nu_s}\varphi \ ,
	\end{split} \label{eq:Fronsdal_00s}
\end{align}
where $\Box$ is the flat d'Alembertian as before, and $C'$ is again an overall coupling constant.

To compare \eqref{eq:lightcone_00s} with \eqref{eq:Fronsdal_00s}, let us embed the lightcone fields $\Phi_0,\Phi_{\pm s}$ into covariant Fronsdal fields in a \emph{lightcone ansatz}:
\begin{align}
	\varphi &= z\Phi_0 \ ; \label{eq:Fronsdal_lightcone_0} \\ 
	\varphi_{\mu_1\dots\mu_s} &= \frac{z^{1-s}}{2^{s/2}(\ell\cdot\del)^s}\,\sigma_{\mu_1}^{\alpha_1\dot\alpha_1}\dots\sigma_{\mu_s}^{\alpha_s\dot\alpha_s}
	    q_{\alpha_1}q^{\beta_1}\dots q_{\alpha_s}q^{\beta_s}  \del_{\beta_1\dot\alpha_1}\dots\del_{\beta_s\dot\alpha_s}\Phi_s + c.c. \ . \label{eq:Fronsdal_lightcone}
\end{align}
In the flat limit (i.e. taking all derivatives as flat), it's easy to see that the ansatz \eqref{eq:Fronsdal_lightcone} is transverse and traceless. Plugging the ansatz \eqref{eq:Fronsdal_lightcone_0}-\eqref{eq:Fronsdal_lightcone} into the kinetic terms of the covariant Lagrangian \eqref{eq:Fronsdal_00s}, we recover the kinetic terms of the lightcone Lagrangian \eqref{eq:lightcone_00s}. Next, we plug \eqref{eq:Fronsdal_lightcone_0}-\eqref{eq:Fronsdal_lightcone} into the cubic vertex of \eqref{eq:Fronsdal_00s}. The result matches with \eqref{eq:lightcone_00s}, if we identify:
\begin{align}
 C = C' \ ; \quad a = 1 \ .
\end{align}
We have thus derived the result \eqref{eq:a}.

\end{document}